\documentclass[aps,pre,10pt,superscriptaddress,footinbib,twocolumn,amsmath,showpacs,amssymb,final,letterpaper]{revtex4-1}
\usepackage{graphicx}
\usepackage[utf8]{inputenc}
\usepackage[colorlinks=true,linkcolor=blue,urlcolor=blue,citecolor=blue,anchorcolor=blue]{hyperref}
\usepackage[export]{adjustbox}% http://ctan.org/pkg/adjustbox
\usepackage{bm}
\hypersetup{pdfdisplaydoctitle=true,pdftitle={Finite-size analysis of the detectability limit of the stochastic block model}}
\hypersetup{pdfauthor={J.-G.  Young  P. Desrosiers  L. Hébert-Dufresne, E. Laurence, L. J. Dubé}}
\hypersetup{pdfsubject={Finite-size analysis of the detectability limit of the stochastic block model}}

\newcommand{\Pf}[2]{\frac{\partial #1}{\partial #2}}
\newcommand{\Pfn}[3]{\frac{\partial^{#3} #1}{\partial #2^{#3}}}
\newcommand{\mean}[1]{\left\langle{#1}\right\rangle}
\newcommand{\avg}[1]{\langle #1 \rangle}
\newcommand{\outs}{\mathrm{out}}
\newcommand{\ins}{\mathrm{in}}
\newcommand{\BigO}{\mathcal{O}}
\newcommand{\T}{\intercal}

\usepackage{amsthm}

\newtheorem{thm}{Theorem}

\usepackage{braket}

\begin{document}
\title{Finite-size analysis of the detectability limit of the stochastic block model}
\date{\today}
\author{Jean-Gabriel Young}
\email{jean-gabriel.young.1@ulaval.ca}
\affiliation{D\'epartement de Physique, de G\'enie Physique, et d'Optique, Universit\'e Laval, Qu\'ebec (Qu{\'e}bec), Canada G1V 0A6}

\author{Patrick Desrosiers}
\affiliation{D\'epartement de Physique, de G\'enie Physique, et d'Optique, Universit\'e Laval, Qu\'ebec (Qu{\'e}bec), Canada G1V 0A6}
\affiliation{Centre de recherche de l'Institut universitaire en sant\'e mentale de Qu\'ebec, Qu\'ebec (Qu\'ebec), Canada G1J 2G3}

\author{Laurent H\'ebert-Dufresne}
 \affiliation{Santa Fe Institute, Santa Fe, New Mexico, USA, 87501}

\author{Edward Laurence}
\affiliation{D\'epartement de Physique, de G\'enie Physique, et d'Optique, Universit\'e Laval, Qu\'ebec (Qu{\'e}bec), Canada G1V 0A6}

\author{Louis J. Dub\'e}
\email{ljd@phy.ulaval.ca}
\affiliation{D\'epartement de Physique, de G\'enie Physique, et d'Optique, Universit\'e Laval, Qu\'ebec (Qu{\'e}bec), Canada G1V 0A6}
\begin{abstract}
It has been shown in recent years that the stochastic block model (SBM) is sometimes undetectable in the sparse limit, i.e., that no algorithm can identify a partition correlated with the partition used to generate an instance, if the instance is sparse enough and infinitely large.
In this contribution, we treat the finite case explicitly, using arguments drawn from information theory and statistics.
We give a necessary condition for finite-size detectability in the general SBM.
We then distinguish the concept of average detectability from the concept of instance-by-instance detectability and give explicit formulas for both definitions.
Using these formulas, we prove that there exist large equivalence classes of parameters, where widely different network ensembles are equally detectable with respect to our definitions of detectability.
In an extensive case study, we investigate the finite-size detectability of a simplified variant of the SBM, which encompasses a number of important models as special cases.
These models include the symmetric SBM, the planted coloring model, and more exotic SBMs not previously studied.
We conclude with three appendices, where we study the interplay of noise and detectability, establish a connection between our information-theoretic approach and random matrix theory, and provide proofs of some of the more technical results.
\end{abstract}

\maketitle

%~~~~~~~~~~~~~~~~~~~~~~~~~~~~~~~~~~~~~~~~~~~~~~~~~~~~~~~~~~~~~~~~~~~~~~~~~~~~~~
\section{Introduction}
\label{section:introduction}
%~~~~~~~~~~~~~~~~~~~~~~~~~~~~~~~~~~~~~~~~~~~~~~~~~~~~~~~~~~~~~~~~~~~~~~~~~~~~~~

Mesoscopic analysis methods \cite{Porter2009} are among the most valuable tools available to applied network scientists and theorists alike.
Their aim is to identify regularities in the structure of complex networks, thereby allowing for a better understanding of their function \cite{Porter2009,Fortunato2010,Newman2012}, their structure \cite{Seshadhri2012,Peixoto2014b}, their evolution \cite{Young2016,HebertDufresne2011}, and of the dynamics they support \cite{Nematzadeh2014,Rosvall2008,HebertDusfresne2016}.
Community detection is perhaps the best-known method of all \cite{Fortunato2010,Porter2009}, but it is certainly not the only one of its kind \cite{Newman2012}.
It has been shown, for example, that the separation of nodes in a core and a periphery occurs in many empirical networks \cite{Borgatti2000}, and that this separation gives rise to more exotic mesoscopic patterns such as overlapping communities \cite{Yang2014}.
This is but an example---there exist multitudes of decompositions in structures other than communities that explain the shape of networks both clearly and succinctly \cite{Peixoto2012}.

The stochastic block model (SBM) has proven to be versatile and principled in uncovering these patterns \cite{Holland1983,Holland1981,White1976}.
According to this simple generative model, the nodes of a network are partitioned in blocks (the \emph{planted partition}), and an edge connects two nodes with a probability that depends on the partition.
The SBM can be used in any of two directions: Either to generate random networks with a planted mesoscopic structure \cite{Nematzadeh2014,HebertDusfresne2016} or to infer the hidden mesoscopic organization of real complex networks, by fitting the model to network datasets  \cite{Holland1983,Snijders1997,Peixoto2012}---perhaps its most useful application.

Stochastic block models offer a number of advantages over other mesoscopic pattern detection methods \cite{Newman2012}.
One, there is no requirement that nodes in a block be densely connected, meaning that blocks are much more general objects than communities.
Two, the sound statistical principles underlying the SBM naturally solve many hard problems that arise in network mesoscopic analysis;
this includes the notoriously challenging problem of determining the optimal number of communities in a network \cite{Peixoto2013,Newman2016b,Kawamoto2016}, or of selecting among the many possible descriptions of a network \cite{Porter2009,Peixoto2015,Kawamoto2016}.

Another advantage of the statistical formulation of the SBM is that one can rigorously investigate its limitations.
It is now known, for example, that the SBM admits a \emph{resolution limit} \cite{Peixoto2013} akin to the limit that arises in modularity--based detection method \cite{Fortunato2007}.
The limitations that have attracted the most attention, however, are the \emph{detectability limit} and the closely related concept of \emph{consistency limit} \cite{Abbe2016b}.
The SBM is said to be detectable for some parameters if an algorithm can construct a partition correlated with the planted partition  \footnote{By \emph{correlated}, it is meant that the two partitions are more similar than two randomly constructed partitions. Our choice of measure will be made explicit at a later stage.},  using no information other than the structure of a single---infinitely large---instance of the model.
It is said to be consistent if one can \emph{exactly} recover the planted partition.
Therefore, consistency begets detectability, but not the other way around.
Understanding when and why consistency (or detectability) can be expected is important, since one cannot trust the partitions extracted by SBM if it operates in a regime where it is not consistent (or detectable) \cite{Abbe2016b}.

Due to rapid developments over the past few years, the locations of the boundaries between the different levels of detectability are now known for multiple variants of the SBM, in the limit of infinite network sizes.
If the average degree scales at least logarithmically with the number of nodes, then the SBM is consistent \cite{Bickel2009,Abbe2016}, unless the constant multiplicative factor is too small, in which case the SBM is then detectable, but not consistent.
If the average degree scales slower than logarithmically, then the SBM is at risk of entering an \emph{undetectable} phase where no information on the planted partition can be recovered from the network structure \cite{Reichardt2008,Decelle2011a}.
This happens if the average degree is a sufficiently small constant independent of the number of nodes.

These asymptotic results are, without a doubt, extremely useful.
Many efficient algorithms have been developed to extract information out of hardly consistent infinite instances \cite{Krzakala2013,Massoulie2014,Decelle2011a,Decelle2011b}.
Striking connections between the SBM and other stochastic processes have been established in the quest to bound the undetectable regime from below \cite{Mossel2013,Banks2016,Abbe2016,Abbe2016b}.
But real networks are not infinite objects.
Thus, even though it has been observed that there is a good agreement between calculations carried out in the infinite-size limit and empirical results obtained on small networks \cite{Decelle2011b}, it is not immediately clear that the phenomenology of the infinite case carries over, unscathed, to the finite case.

In this paper, we explicitly investigate detectability in \emph{finite} networks generated by the SBM.
We understand detectability in the information-theoretic sense \cite{Banks2016}; our analysis is therefore algorithm--independent, and yields the boundaries of the region of the parameter space where the planted partition is undetectable, even for an optimal algorithm (with possibly exponential running time).

The combination of this information-theoretic point of view with our finite-size analysis leads to new insights and results, which we organize as follows.
We begin by formally introducing the SBM and the necessary background in Sec.~\ref{section:stochastic_block_model}.
We use this section to briefly review important notions, including inference (Sec.~\ref{subsection:stochastic_block_model-inference}), as well as the consistency and detectability of the infinite SBM (Sec.~\ref{subsection:stochastic_block_model-related_work}).
In Sec.~\ref{section:finite_limit}, we present a necessary condition for detectability, and show that it is always met, on average, by finite instances of the SBM.
We then establish the existence of a large equivalence class with respect to this notion of average detectability.
In Sec.~\ref{section:eta-detectability}, we introduce the related concept of $\eta$--detectability and investigate the complete detectability distribution, beyond its average.
In Sec.~\ref{section:case_study}, we apply the perfectly general framework of Secs.~\ref{section:finite_limit}--\ref{section:eta-detectability} to a constrained variant of the SBM: the general modular graph model of Ref.~\cite{Kawamoto2017}.
The results of this section hold for a broad range of models, since the general modular graphs encompass the symmetric SBM, the planted coloring model, and many other models as special cases.
We gather concluding remarks and open problems in Sec.~\ref{section:discussion}.
Three appendices follow.
In the first, we investigate the interplay between noise and our notion of average detectability (Appendix~\ref{appendix:noisy_sbm}); in the second, we establish a connection between our framework and random matrix theory (Appendix~\ref{appendix:connection_with_rmt}); in the third, we give the details of two technical proofs encountered in the main text (Appendix~\ref{appendix:proof_details}).

%~~~~~~~~~~~~~~~~~~~~~~~~~~~~~~~~~~~~~~~~~~~~~~~~~~~~~~~~~~~~~~~~~~~~~~~~~~~~~~
\section{Stochastic block model}
\label{section:stochastic_block_model}
%~~~~~~~~~~~~~~~~~~~~~~~~~~~~~~~~~~~~~~~~~~~~~~~~~~~~~~~~~~~~~~~~~~~~~~~~~~~~~~

%~~~~~~~~~~~~~~~~~~~~~~~~~~~~~~~~~~~~~~~~~~~~~~~~~~~~~~~~~~~~~~~~~~~~~~~~~~~~~~
\subsection{Definition of the model}
\label{subsection:stochastic_block_model-definition}
%~~~~~~~~~~~~~~~~~~~~~~~~~~~~~~~~~~~~~~~~~~~~~~~~~~~~~~~~~~~~~~~~~~~~~~~~~~~~~~
The stochastic block model is formally defined as follows:
Begin by partitioning a set of $n$ nodes  in $q$ blocks of fixed sizes $\bm{n}=(n_1,...,n_q)$, with $n=\sum_{r=1}^q n_r$.
Denote this partition by $\mathcal{B}=\{B_1,...,B_q\}$, where $B_r$ is the set of nodes in the $r$\textsuperscript{th} block.
Then, connect the nodes in block $B_r$ to the nodes in block $B_s$ with probability $p_{rs}$.
In other words, for each pair of nodes $(v_i,v_j)$, set the element $a_{ij}$ of the adjacency matrix $\bm{A}$ to 1 with probability $p_{\sigma(v_i)\sigma(v_j)}$ and to 0 otherwise, where $\sigma(v_i)$ is the block of $v_i$.
Note that for the sake of clarity, we will obtain all of our results for simple graphs, where edges are undirected and self-loops (edges connecting a node to itself) are forbidden 
\footnote{There is no obstacle to a generalization to the directed case (with or without self-loops).}.
This implies that $p_{rs}=p_{sr}$ and that $a_{ii}=0$.

We will think of this process as determining the outcome of a random variable, whose support is the set of all networks of $n$ nodes.
Due to the independence of edges, the probability (likelihood) of generating a particular network $G$ is simply given by the product of $\binom{n}{2}$ Bernoulli random variables, i.e.,
\begin{equation}
    \label{eq:likelihood_aij}
    \mathbb{P}(G|\mathcal{B}, \bm{P})  = \prod_{i < j} [1- p_{\sigma(v_i)\sigma(v_j)}]^{1-a_{ij}}[p_{\sigma(v_i)\sigma(v_j)}]^{a_{ij}}\;,
\end{equation}
where $\bm{P}$ is the $q\times q$ matrix of connection probabilities of element $p_{rs}$ (sometimes called the affinity or density matrix), and $i < j$ is a shorthand for ``$i,j:\ 1\leq i < j \leq n$.''
It is easy to check that the probability in Eq.~\eqref{eq:likelihood_aij} is properly normalized over the set of all networks of $n$ distinguishable nodes.

A useful alternative to Eq.~\eqref{eq:likelihood_aij} expresses the likelihood  in terms of the number of edges between each pair of blocks $(B_r, B_s)$ rather than as a function of the adjacency matrix \cite{Snijders1997}.
Notice how the number of edges $m_{rs}$ appearing between the sets of nodes $B_r$ and $B_s$ is at most equal to
\begin{equation}
    \label{eq:edge_counts_undirected}
    m_{rs}^{\max} = \left\{\begin{array}{ll} \binom{n_r}{2}&\text{if $r=s$,}\\ n_rn_s& \text{otherwise.} \end{array}\right.
\end{equation}
Each of these $m^{\max}_{rs}$ edges exists with probability $p_{rs}$.
This implies that $m_{rs}$ is determined by the sum of $m_{rs}^{\max}$ Bernoulli trials of probability $p_{rs}$, i.e., that $m_{rs}$ is a binomial variable of parameter $p_{rs}$ and maximum $m_{rs}^{\max}$.
The probability of generating a particular instance $G$ can therefore be written equivalently as
\begin{equation}
    \label{eq:likelihood_edges}
    \mathbb{P}(G|\mathcal{B}, \bm{P})  = \prod_{r\leq s} (1- p_{rs})^{m_{rs}^{\max}-m_{rs}}(p_{rs})^{m_{rs}}\;.
\end{equation}
where $\{m_{rs}\}$ and $\{m_{rs}^{\max}\}$ are jointly determined by the partition $\mathcal{B}$ and the structure of $G$, and $r\leq s$ denotes ``$r,s:\ 1\leq r \leq s \leq q$.''

Having a distribution over all networks of $n$ nodes, one can then compute average values over the ensemble.
For example, the average degree of node $v_i$ is given by
\begin{equation}
    \avg{k_i} = \sum_{r}p_{\sigma(v_i)r}(n_r - \delta_{\sigma(v_i)r})\;,
\end{equation}
where $\delta_{ij}$ is the Kronecker Delta.
The expression correctly depends on the block $B_\sigma(v_i)$ of $v_i$; nodes in different blocks will, in general, have different average degree.
Averaging over all nodes, one finds the average degree of the network
\begin{equation}
    \avg{k} = \frac{2}{n}\sum_{r\leq s} m_{rs}^{\max}p_{rs}\;.
\end{equation}
This global quantity determines the density of the SBM when $n\to\infty$.
The SBM is said to be dense if $\avg{k} =\BigO(n)$, i.e., if $p_{rs}$ is a constant independent of $n$.
It is said to be sparse if $\avg{k} = \BigO(1)$, i.e., if $p_{rs}=c_{rs}/n$ goes to zero as $n^{-1}$.
In the latter case, a node has a constant number of connections even in an infinitely large network---a feature found in most large scale real networks \cite{Newman2010}.

For finite instances, it will often be more useful to consider the average density directly.
It is defined as the number of edges in $G$, normalized by the number of possible edges, i.e.,
\begin{equation}
    \label{eq:density_def}
    \rho = \frac{\avg{k}}{n-1} = \sum_{r\leq s} (m^{\max}_{rs}/m^{\max})p_{rs} \equiv \sum_{r\leq s} \alpha_{rs}p_{rs}\;,
\end{equation}
where $m^{\max}=\sum_{r\leq s}m_{rs}^{\max}$, and 
\begin{equation}
    \label{eq:def_alpha}
    \alpha_{rs} := m_{rs}^{\max}/m^{\max}\;.
\end{equation}
The dense versus sparse terminology is then clearer: The density of sparse networks goes to zero as $\BigO(n^{-1})$, while dense networks have a nonvanishing density $\rho=\BigO(1)$.

%~~~~~~~~~~~~~~~~~~~~~~~~~~~~~~~~~~~~~~~~~~~~~~~~~~~~~~~~~~~~~~~~~~~~~~~~~~~~~~
\subsection{Inference}
\label{subsection:stochastic_block_model-inference}
%~~~~~~~~~~~~~~~~~~~~~~~~~~~~~~~~~~~~~~~~~~~~~~~~~~~~~~~~~~~~~~~~~~~~~~~~~~~~~~

Depending on the elements of $\bm{P}$, the SBM can generate instances reminiscent of real networks with, e.g., a community structure \cite{Newman2012} ($p_{rr}>p_{rs}$) or a core-periphery organization \cite{Borgatti2000} ($p_{11}>p_{12}>p_{22}$ and $p_{22}\sim0$).
However, the SBM really shines when it is used to infer the organization in blocks of the nodes of real complex networks---this was, after all, its original purpose \cite{Holland1983}. 

To have inferred the mesoscopic structure of a network (with the SBM) essentially means that one has found the partition $\mathcal{B}^*$ and density matrix $\bm{P}^*$ that best describes it.
In principle, it is a straightforward task, since one merely needs to (a) assign a likelihood $\mathbb{P}(\mathcal{B},\bm{P}|G)$ to each pair of  partition and parameters [see Eqs.~\eqref{eq:likelihood_aij}--\eqref{eq:likelihood_edges}], then (b) search for the most likely pair ($\mathcal{B}^*$, $\bm{P}^*$).
Since there are exponentially many possible partitions, this sort of enumerative approach is of little practical use.
Fortunately, multiple approximate and efficient inference tools have been proposed to circumvent this fundamental problem.
They draw on ideas from various fields such as statistical physics \cite{Decelle2011a,Decelle2011b,Peixoto2012}, Bayesian statistics \cite{Snijders1997,Van2016}, spectral theory \cite{Krzakala2013,Massoulie2014,Newman2016,Newman2013}, and graph theory \cite{Condon2001}, to name a few, and they all produce accurate results in general.

%~~~~~~~~~~~~~~~~~~~~~~~~~~~~~~~~~~~~~~~~~~~~~~~~~~~~~~~~~~~~~~~~~~~~~~~~~~~~~~
\subsection{Detectability and consistency}
\label{subsection:stochastic_block_model-related_work}
%~~~~~~~~~~~~~~~~~~~~~~~~~~~~~~~~~~~~~~~~~~~~~~~~~~~~~~~~~~~~~~~~~~~~~~~~~~~~~~
One could expect perfect recovery of the parameters and partition from most of these sophisticated algorithms.
This is called the consistency property.
It turns out, however, that all known inference algorithms for the SBM, as diverse as they might be, fail on this account.
And their designs are not at fault, for there exists an explanation of this generalized failure.

Consider the density matrix of elements $p_{rs}=\rho$ $\forall (r,s)$.
It is clear that the block partition is irrelevant---the generated network cannot and will not encode the planted partition.
Thus, no algorithm will be abe to differentiate the planted partition from other partitions.
It is then natural to assume that inference will be hard or impossible if $p_{rs}=\rho + \epsilon_{rs}(n)$, where $\epsilon_{rs}(n)$ is a very small perturbation for networks of $n$ nodes; there is little difference between the uniform case and this perturbed case.
In contrast, if the elements of $\bm{P}$ are widely different from one another, e.g., if $p_{rr}=1$ and $p_{rs}=0$ for $r \neq s$, then easy recovery should be expected.

Understanding where lies the transition between these qualitatively different regimes has been the subject of much recent research  (see Ref.~\cite{Abbe2016b} for a recent survey).
As a result, the regimes have been clearly separated as follows: (i) the undetectable regime, (ii) the detectable (but not consistent) regime and (iii) the consistent regime (and detectable).
It has further been established that the scaling of $\rho$ with respect to $n$ determines which regime is reached, in the limit $n\to\infty$.

The SBM is said to be \emph{strongly consistent} if its planted partition can be inferred perfectly, with a probability that goes to 1 as $n\to\infty$ (it is also said to be in the \emph{exact recovery} phase).
Another close but weaker definition of consistency asks that the probability of misclassifying a node goes to zero with $n\to\infty$ (the \emph{weakly consistent} or \emph{almost exact recovery} phase).
These regimes prevail when $\bm{P}$ scales at least as fast as $\bm{P}=\log(n)\bm{C}/n$, where $\bm{C}$ is a $q\times q$ matrix of constants \cite{Abbe2015,Abbe2016,Bickel2009}.
Predictably, most algorithms (e.g., those of Refs.~\cite{Condon2001,Snijders1997,Abbe2015}) work well in the exact recovery phase regime, since it is the easiest of all .

In the \emph{detectable} (but not consistent) regime, exact recovery is no longer possible  (the \emph{partial recovery} phase).
The reason is simple: through random fluctuations, some nodes that belong to, say, block $B_1$, end up  connecting to other nodes as if they belonged to block $B_2$.
They are thus systematically misclassified, no matter the choice of algorithms.
This occurs whenever $\bm{P}=\bm{C}/n$, or $\bm{P}=f(n)\bm{C}/n$, with $f(n)$ a function of $n$ that scales slower than $\log(n)$.

The discovery of the third regime---the \emph{undetectable regime}---arguably rekindled the study of the fundamental limits of the SBM \cite{Reichardt2008,Decelle2011a}.
In this regime, which occurs when $\bm{P}=\bm{C}/n$ and $\bm{C}$ is more or less uniform, it is impossible to detect a partition that is even correlated with the planted one.
That is, one cannot classify nodes better than at random, and no information on the planted partition can be extracted.
Thus, some parametrizations of the SBM are said to lie below the \emph{detectability limit}.
This limit was first investigated with informal arguments from statistical physics \cite{Reichardt2008,Decelle2011a,Decelle2011b,Kawamoto2017,Zhang2014}, and has since been rigorously formalized in Refs.~\cite{Mossel2014,Banks2016}, among others.

There exist many efficient algorithms that are reliable close to the detectability limit; noteworthy examples include belief propagation \cite{Decelle2011a,Decelle2011b,Zhang2015c}, and spectral algorithms based on the ordinary \cite{Krzakala2013} and weighted \cite{Mossel2013} non backtracking matrix, as well as matrices of self-avoiding walks \cite{Massoulie2014}.
But when the number of blocks is too large, most of these algorithms are known to fail well above the information-theoretic threshold, i.e., the point where it can be proven that the partition is detectable given arbitrary computational power.
It has been therefore conjectured in Ref.~\cite{Decelle2011b}, that there exists multiple troublesome phases for inference: A truly undetectable regime, and a regime where detection is not achievable \emph{efficiently}.
In the latter, it is thought that one \emph{can} find a good partition, but only by enumerating all partitions---a task of exponential complexity.

In this contribution, however, we will not focus on this so-called hard regime.
As far as we are concerned, detectability will be understood in terms of information, i.e.,  we will delimit the boundaries of the information-theoretically undetectable regime.

%~~~~~~~~~~~~~~~~~~~~~~~~~~~~~~~~~~~~~~~~~~~~~~~~~~~~~~~~~~~~~~~~~~~~~~~~~~~~~~
\section{Detectability of finite networks}
\label{section:finite_limit}
%~~~~~~~~~~~~~~~~~~~~~~~~~~~~~~~~~~~~~~~~~~~~~~~~~~~~~~~~~~~~~~~~~~~~~~~~~~~~~~
Detectability and consistency are well-separated phases of the infinite stochastic block model.
A minute perturbation to the parameters may potentially translate into widely different qualitative behaviors.
The picture changes completely when one turns to finite instances of the model.
Random fluctuations are not smoothed out by limits, and transitions are much less abrupt.
We argue that, as a result, one has to account for the complete distribution of networks to properly quantify detectability, i.e., define detectability for \emph{network instances} rather than parameters.
This, in turn, commands a different approach that we now introduce.

%~~~~~~~~~~~~~~~~~~~~~~~~~~~~~~~~~~~~~~~~~~~~~~~~~~~~~~~~~~~~~~~~~~~~~~~~~~~~~~
\subsection{Hypothesis test and the detectability limit}
\label{subsection:finite_limit-hypotheses_testing}
%~~~~~~~~~~~~~~~~~~~~~~~~~~~~~~~~~~~~~~~~~~~~~~~~~~~~~~~~~~~~~~~~~~~~~~~~~~~~~~
Consider a single network $G$, generated by the SBM with some planted partition $\mathcal{B}$ and matrix $\bm{P}=r\bm{11}^\T+\bm{\epsilon}$, where $\bm{11}^\T$ is a matrix of ones, $r$ a constant, and $\bm{\epsilon}$ a matrix of (small) fluctuations.
Suppose that the average density equals $\rho$, and consider a second density matrix $\rho\bm{11}^\T$ for which the block structure has no effect on the generative process.
If an observer with \emph{complete knowledge} of the generative process and its parameters cannot tell which density matrix, $\bm{P}$ or $\rho\bm{11}^\T$, is the most likely to have generated $G$, then it is clear that \emph{this particular instance} does not encode the planted partition.
As a result, it will be impossible to detect a partition correlated with the planted partition.

This idea can be translated into a mathematical statement by way of a likelihood test.
For a SBM of average density $\rho$, call the ensemble of Erd\H{o}s-R\'enyi graphs of density $\rho$ the ensemble of \emph{equivalent random networks}.
Much like the SBM (see Sec.~\ref{section:stochastic_block_model}), its likelihood $Q(G|\rho)$ is given by the product of the density of $\binom{n}{2}$ independent and identically distributed  Bernoulli variables, i.e.,
\begin{equation}
    \label{eq:likelihood_erg}
    \mathbb{Q}(G|\rho) = \prod_{i<j} \rho^{a_{ij}}(1-\rho)^{a_{ij}}=\rho^{m}(1-\rho)^{m^{\max}-m}\;,
\end{equation}
where $m:=\sum_{r\leq s} m_{rs}$ is the total number of edges in $G$.

The condition is then the following: Given a network $G$ generated by the SBM of average density $\rho$ and density matrix $\bm{P}$, one can detect the planted partition $\mathcal{B}$ if the SBM is more likely than its equivalent random ensemble of density $\rho$, i.e.,
\begin{equation}
    \label{eq:ratio_randomvariable}
    \Lambda = \frac{\mathbb{P}(G|\mathcal{B},\bm{P})}{\mathbb{Q}(G|\rho)} > 1\;.
\end{equation}
A similar condition has been used in Refs.~\cite{Mossel2014} and \cite{Banks2016}  to pinpoint the location of the detectability limit in infinite and sparse instances of the SBM.
Nothing forbids its application to the finite-size problem;
we will see shortly that it serves us well in the context of finite-size detectability.

%~~~~~~~~~~~~~~~~~~~~~~~~~~~~~~~~~~~~~~~~~~~~~~~~~~~~~~~~~~~~~~~~~~~~~~~~~~~~~~
\subsection{Normalized log-likelihood ratio}
\label{subsection:normalized_log_likelihood}
%~~~~~~~~~~~~~~~~~~~~~~~~~~~~~~~~~~~~~~~~~~~~~~~~~~~~~~~~~~~~~~~~~~~~~~~~~~~~~~
The (equivalent) normalized log-likelihood ratio
\begin{equation}
    \mathcal{L} := \frac{\log\Lambda}{m^{\max}}
\end{equation}
will be more practical for our purpose.
This simple transformation brings the line of separation between models from $\Lambda = 1$ to $\mathcal{L} = 0$, and prevents the resulting quantity from becoming too large.
More importantly, it changes products into sums and allows for a simpler expression,
\begin{equation}
     \mathcal{L} \!=\!\! 
     \sum_{r\leq s}\Bigg\{\!\!\frac{m_{rs}}{m^{\max}} \log\!\!\left[\frac{p_{rs}(1-\rho)}{\rho(1-p_{rs})}\right] +\alpha_{rs}\!\log\!\!\left[\frac{1-p_{rs}}{1-\rho}\right]\!\!\Bigg\}.
     \label{eq:normalized_log_likelihood}
\end{equation}

\begin{figure} 
    \centering
    \quad\includegraphics[width=0.95\linewidth]{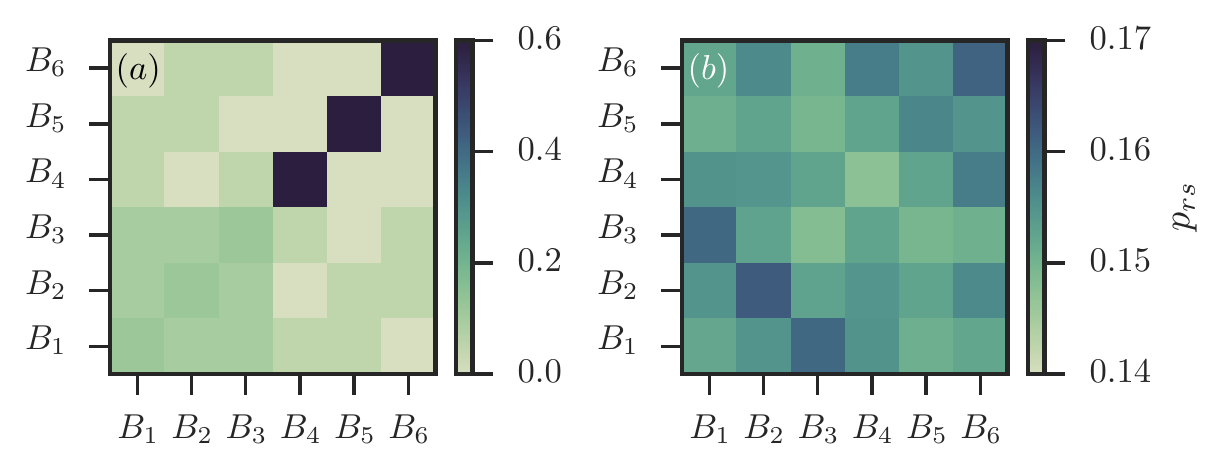} 
    \includegraphics[width=\linewidth, trim=1cm 0cm 0cm 0cm, clip=true]{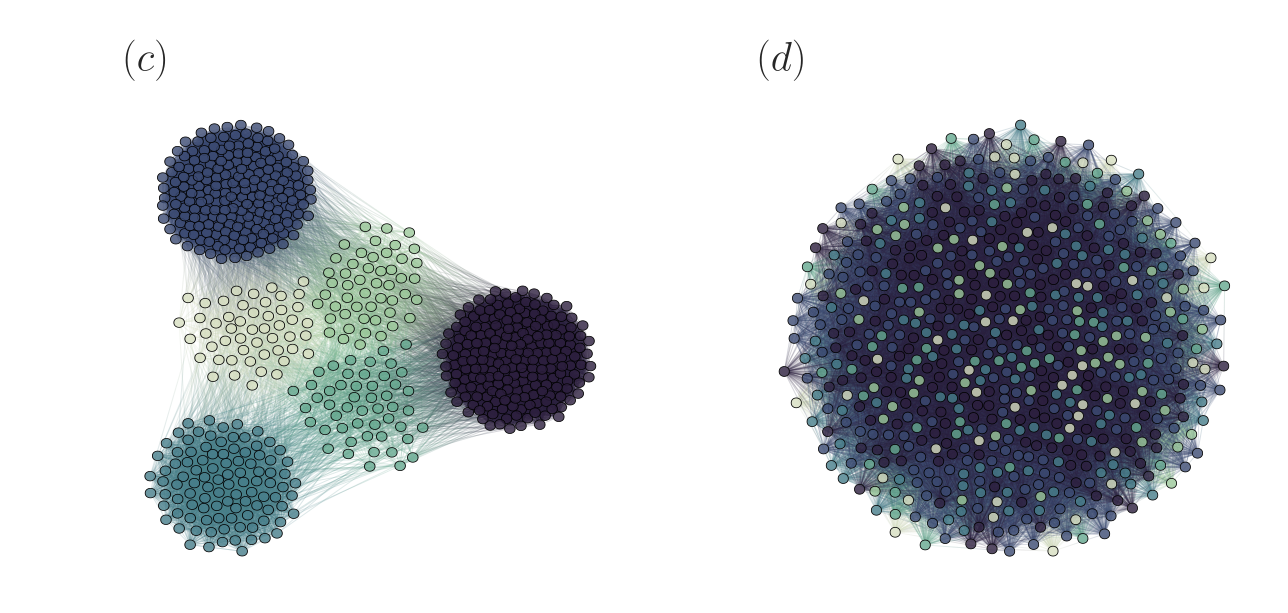}\\
    \includegraphics[width=0.99\linewidth]{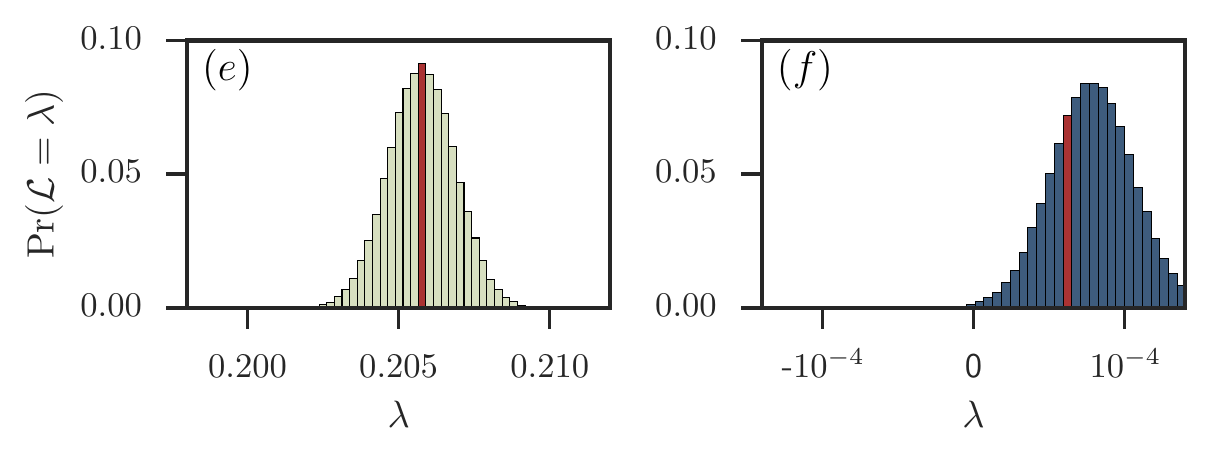} 
    \caption{Stochastic block model with (a, c, e) non uniform density matrix and (b, d, f) nearly uniform density matrix.
    (a, b) Density matrix of the two ensembles. Notice the difference in scale.
    (c, d) One instance of each ensemble, with $\bm{n}=[50,50,50,100,200,200]$.
    Each color denotes a block \cite{graph_tool}.
    (e, f) Empirical distribution of the normalized log-likelihood obtained from $100\;000$ samples of $\mathcal{L}$.
    The bins in which the instances (c, d) fall are colored in red.
    Notice that a negative log-likelihood ratio is associated with some instances in (f).
    }
    \label{fig:log_likelihood_full_distribution} 
\end{figure}

We will focus, for the remainder of this paper, on the case where network instances $G$ of $n$ nodes are drawn from the SBM of parameters $(\mathcal{B},\bm{P})$.
In this context, $\mathcal{L}$ can is a random variable whose support is the networks of $n$ nodes with labeled nodes (see Fig.~\ref{fig:log_likelihood_full_distribution}).
Since $\bm{P},\rho,\alpha,$ and $m^{\max}$ are all parameters, $\mathcal{L}$ can also be seen as a weighted sum of binomial distributed random variables $m_{rs} \sim \mathrm{Bin}(m^{\max}_{rs} , p_{rs})$, with a constant offset.
Its average  will be a prediction of the detectability for the ensemble (Sec.~\ref{section:average_detectability}), and the probability $\Pr(\mathcal{L}<0;\bm{P},\bm{\alpha}, m^{\max})$ will give the fraction of instances that are  undetectable for the selected parameters  (Sec.~\ref{section:eta-detectability}).

%~~~~~~~~~~~~~~~~~~~~~~~~~~~~~~~~~~~~~~~~~~~~~~~~~~~~~~~~~~~~~~~~~~~~~~~~~~~~~~
\subsection{Interpretation of $\mathcal{L}$: Information-theoretic bound and inference difficulty}
%~~~~~~~~~~~~~~~~~~~~~~~~~~~~~~~~~~~~~~~~~~~~~~~~~~~~~~~~~~~~~~~~~~~~~~~~~~~~~~
Because likelihood ratio tests can be understood as quantifying the amount of evidence for a hypothesis (compared to a null hypothesis), there will be two interpretations of $\mathcal{L}$.

On the one hand, the condition $\mathcal{L}>0$ will provide a lower bound on detectability; if $\mathcal{L}(G,\mathcal{B},\bm{P})<0$, then we can safely say that the instance $G$ is information-theoretically undetectable.
However, $\mathcal{L}(G,\mathcal{B},\bm{P})>0$ does not necessarily mean that the instance is information-theoretically detectable.
This is due to the fact that the condition $\mathcal{L}>0$ is necessary but not sufficient, since we assume a complete knowledge of the generative process in calculating $\mathcal{L}$.

On the other hand, we will interpret $\mathcal{L}$ operationally as a measure of the difficulty of the inference problem (not in the computational sense).
A large ratio of a hypothesis $\mathbb{H}$ to its null model $\mathbb{H}_0$ implies that the hypothesis is a much better explanation of the data than $\mathbb{H}_0$;
therefore, $\mathcal{L}$ measures how easy it is to select between $\mathbb{P}$ and $\mathbb{Q}$, given full knowledge of the generative process, and inference algorithms will perform better when the ratio is larger.
Many empirical results will validate this interpretation (see Sec.~\ref{section:case_study}).

%~~~~~~~~~~~~~~~~~~~~~~~~~~~~~~~~~~~~~~~~~~~~~~~~~~~~~~~~~~~~~~~~~~~~~~~~~~~~~~
\section{Average detectability}
\label{section:average_detectability}
%~~~~~~~~~~~~~~~~~~~~~~~~~~~~~~~~~~~~~~~~~~~~~~~~~~~~~~~~~~~~~~~~~~~~~~~~~~~~~~

%~~~~~~~~~~~~~~~~~~~~~~~~~~~~~~~~~~~~~~~~~~~~~~~~~~~~~~~~~~~~~~~~~~~~~~~~~~~~~~
\subsection{Average normalized log-likelihood}
\label{subsection:average_of_normalized_loglikelihood}
%~~~~~~~~~~~~~~~~~~~~~~~~~~~~~~~~~~~~~~~~~~~~~~~~~~~~~~~~~~~~~~~~~~~~~~~~~~~~~~
The average of a log-likelihood ratio is also known as the Kullback-Leibler (KL) divergence $D(\cdot||\cdot)$ of two hypotheses \cite{Cover2012}, i.e.,
\begin{align}
    \label{eq:KL_divergence}
    \avg{\mathcal{L}(\bm{\alpha},\bm{P})} &=  \sum_{\{G\}} \frac{\mathbb{P}(G|\mathcal{B},\bm{P})}{m^{\max}} \log\frac{\mathbb{P}(G|\mathcal{B},\bm{P})}{\mathbb{Q}(G|\rho)} \notag\\
    &= \frac{D(\mathbb{P}||\mathbb{Q})}{m^{\max}}\;,
\end{align}
where the sum runs over all $n$ nodes networks.
Since the KL divergence is always greater or equal to zero, with equality if and only if $\mathbb{P}=\mathbb{Q}$, and since $\mathcal{L}>0$ is only a necessary condition for detectability, the average $\avg{\mathcal{L}}$ will not be enough to conclude on detectability of the SBM, except for the case $\mathbb{P}=\mathbb{Q}$ \footnote{$D(\mathbb{P}||\mathbb{Q})$ also goes to 0 at $\rho=0$, and a more careful scaling analysis is necessary to conclude on the detectability of sparse instances.}.
Results pertaining to $\avg{\mathcal{L}}$  will therefore be best interpreted in terms of inference difficulty.

However, even if the average log-likelihood ratio is always positive (assuming $\mathbb{P}\neq\mathbb{Q}$), it can be extremely close to zero for density matrix $\bm{P}$ ``close'' to $\rho\bm{1}\bm{1}^\T$ [Fig.~\ref{fig:log_likelihood_full_distribution} (f)].
In fact, as we will see in Sec.~\ref{section:eta-detectability}, $\avg{\mathcal{L}(\bm{\alpha},\bm{P})} \approx 0$ implies that there are instances for which $\mathcal{L}<0$.
Therefore, whenever the average is small, we may also take it as a sign that the planted partition of some instances are truly undetectable.

%~~~~~~~~~~~~~~~~~~~~~~~~~~~~~~~~~~~~~~~~~~~~~~~~~~~~~~~~~~~~~~~~~~~~~~~~~~~~~~
\subsection{Compact form}
%~~~~~~~~~~~~~~~~~~~~~~~~~~~~~~~~~~~~~~~~~~~~~~~~~~~~~~~~~~~~~~~~~~~~~~~~~~~~~~
While Eq.~\eqref{eq:KL_divergence} has a precise information-theoretic interpretation, there exists an equivalent form, both more compact and easier to handle analytically.
It is given by
\begin{equation}
    \avg{\mathcal{L}(\bm{\alpha},\bm{P})} = h(\rho) - \sum_{r\leq s}\alpha_{
    rs} h(p_{rs})\;, \label{eq:avg_log_likelihood_entropy_difference}
\end{equation}
where 
\begin{equation}
    h(p)=-(1-p)\log(1-p) - p\log (p)
\end{equation}
is the binary entropy of $p\in[0,1]$. 
This expression can be obtained in a number of ways, the most direct of which is to take the average of Eq.~\eqref{eq:normalized_log_likelihood} over all symmetric matrices $\bm{m}=(m_{11},m_{12},\hdots,m_{qq})$ with entries in $\mathbb{N}$ and upper bounds given by $\bm{m}^{\max}=(m_{11}^{\max},m_{12}^{\max},\hdots,m_{qq}^{\max})$.
That is to say, we use the interpretation where $\mathcal{L}$ is a weighted sum of binomial distributed random variable, instead of the interpretation where it is a random variable over the networks of $n$ nodes (see Sec.~\ref{subsection:normalized_log_likelihood}).
The probability mass function associated to $\bm{m}$ is then $\Pr[\bm{m}] = \prod_{r\leq s} \Pr[m_{rs}]$, where $\Pr[m_{rs}]$ is the binomial distribution of parameter $p_{rs}$ and upper bound $m_{rs}^{\max}$.
Due to the linearity of expectations, it is straightforward to check that the average of the first sum of Eq.~\eqref{eq:normalized_log_likelihood} equals
\begin{multline*}
    \sum_{\bm{m}} \Pr[{\bm{m}}]  \sum_{r\leq s} \frac{m_{rs}}{m^{\max}}  \log\left[\frac{p_{rs}}{\rho}\frac{1-\rho}{1-p_{rs}}\right]\\=\sum_{r\leq s} \log\left[\frac{p_{rs}}{\rho}\frac{1-\rho}{1-p_{rs}}\right] \frac{m^{\max}_{rs}p_{rs}}{m^{\max}}\;.
\end{multline*}
Recalling Eq.~\eqref{eq:density_def}, one then finds
\begin{align*}
   \avg{\mathcal{L}(\bm{\alpha},\bm{P})}
   &=  - \sum_{r\leq s} \alpha_{rs} \bigl[(1-p_{rs})\log(1-\rho) \!+\! p_{rs}\log \rho\bigr]\notag\\
   &\quad + \sum_{r\leq s} \alpha_{rs} [(1-p_{rs})\log(1-p_{rs}) \!+\! p_{rs}\log p_{rs}]\notag\\
   &= h(\rho) -\sum_{r\leq s} \alpha_{rs}\, h(p_{rs}) \;.\label{eq:average_likelihood}
\end{align*}
where $\alpha_{rs}$ is defined in Eq.~\eqref{eq:def_alpha} with the normalization $\sum_{r\leq s} \alpha_{rs} = 1$.
Notice how this expression does not depend on $\mathcal{B}$ anymore.
In this context, the only role of the planted partition is to fix the relative block sizes $\bm{\alpha}$.
Thus, the average log-likelihood $\avg{\mathcal{L}}$ of two models with different planted partitions but identical $\bm{\alpha}$ is the same (up to a size-dependent constant).

With these two expressions for $\avg{\mathcal{L}}$ in hand [Eqs.~\eqref{eq:KL_divergence} and \eqref{eq:avg_log_likelihood_entropy_difference}], we can now build an intuition for what the easiest and most difficult detectability problems might look like.
The KL divergence is never negative, and Eq.~\eqref{eq:avg_log_likelihood_entropy_difference} shows that the maximum of $\avg{\mathcal{L}}$ is $h(1/2)$;
the average of the normalized log-likelihood is thus confined to the interval 
\begin{equation}
    0\leq \avg{\mathcal{L}(\bm{\alpha},\bm{P})} \leq h(1/2)\;.
\end{equation}
An example of parameters that achieves the upper bound would be the SBM of density matrix $p_{11}=p_{22}=1$, $p_{12}=0$, with $\bm{n}=[n/2,n/2]$, i.e., the ``ensemble'' of disconnected $n/2$--cliques (which contains a single instance).
An example of parameters that achieves the lower bound would be $\mathbb{P}=\mathbb{Q}$, but also $\rho\to 0$ [see Eq.~\eqref{eq:avg_log_likelihood_entropy_difference}].

%~~~~~~~~~~~~~~~~~~~~~~~~~~~~~~~~~~~~~~~~~~~~~~~~~~~~~~~~~~~~~~~~~~~~~~~~~~~~~~
\subsection{Equivalent stochastic block models}
\label{subsection:equivalence_classes_avg}
%~~~~~~~~~~~~~~~~~~~~~~~~~~~~~~~~~~~~~~~~~~~~~~~~~~~~~~~~~~~~~~~~~~~~~~~~~~~~~~
We now use Eq.~\eqref{eq:avg_log_likelihood_entropy_difference} to uncover hidden connections between different regimes of the SBM.
Notice how this expression induces equivalence classes in the parameter space of the model, with respect to  $\avg{\mathcal{L}}$, i.e., subsets of parameters that all satisfy
\begin{equation}
    \label{eq:hypersurface_equation}
    \lambda = \avg{\mathcal{L}(\bm{P}, \bm{\alpha})}\;,
\end{equation}
where $\lambda$ is a constant that defines the equivalence class.

In the next paragraphs, we will characterize these equivalence classes in two complementary ways.
First, we will look for global transformations that preserve $\lambda$ and map parameters $(\bm{\alpha},\bm{P})$ to some other---not necessarily close---pair of parameters $(\bm{\alpha}',\bm{P}')$.
Provided that they satisfy a number of standard constraints, these transformations will be shown to correspond to the symmetry group of the set of \emph{hypersurfaces} $\avg{\mathcal{L}(\bm{\alpha},\bm{P})}=\lambda$.
Second, we will consider Eq.~\eqref{eq:hypersurface_equation} explicitly and use it to obtain an approximate hypersurface equation.
This equation will be used in later sections to determine the location of the hypersurfaces that separate the parameter space of the SBM in different detectability phases.

%~~~~~~~~~~~~~~~~~~~~~~~~~~~~~~~~~~~~~~~~~~~~~~~~~~~~~~~~~~~~~~~~~~~~~~~~~~~~~~
\subsubsection{Global transformations: The symmetry group of the SBM}
\label{subsubsection:symmetries_avg_detectability}
%~~~~~~~~~~~~~~~~~~~~~~~~~~~~~~~~~~~~~~~~~~~~~~~~~~~~~~~~~~~~~~~~~~~~~~~~~~~~~~

We first look for the set of $\lambda$--preserving global transformations, i.e., all transformations $T(f_1,f_2)$ of the form 
\begin{equation}
    \label{eq:global_transformation_form}
    \bm{\alpha}' = f_1(\bm{\alpha}),\quad \bm{P}' = f_2(\bm{P})\;
\end{equation}
valid at every point of the parameter space.
This is a broad definition and it must be restricted if we are to get anything useful out of it.
Intuitively, we do not want these transformations to change the space on which they operate, so it is natural to ask that they be space-preserving.
Under the (reasonable) constraint that these transformations are invertible as well, we can limit our search for $\lambda$--preserving transformations to the symmetry group of the parameter space.

We will be able to harness known results of geometry and algebra once the parameter space of the SBM is properly defined.
This space is in fact the Cartesian product of two parameter spaces: The parameter space of $\bm{\alpha}$ and that of $\bm{P}$.
Since there is $q^*=q(q+1)/2$ free parameters in both $\bm{\alpha}$ and $\bm{P}$, the complete space is of dimension  $2q^*-1$.
It is the product of the $q^*$--dimensional hypercube---in which every point corresponds to a choice of $\bm{P}$---, and the $(q^*-1)$--dimensional simplex---in which every point corresponds to a choice of $\bm{\alpha}$.
The latter is a simplex due to the normalization $\sum_{r\leq s}\alpha_{rs}=(m^{\max})^{-1}\sum_{r\leq s}m_{rs}^{\max}=1$.

Now, the symmetry group of the $q^*$--dimensional hypercube and that of the $(q^*-1)$--dimensional regular simplex are well-known \cite{Coxeter1973}: They are respectively the hyperoctahedral group $B_{q^*}$ and the symmetric group $S_{q^*}$.
Their action on $\bm{\alpha}$ and $\bm{P}$ can be described as
\begin{align*} 
    \alpha_{rs}&\mapsto \alpha_{rs}' = \alpha_{\pi(r,s)}\;,\\
    p_{rs}&\mapsto p_{rs}' = \gamma_{rs} + (1 - 2\gamma_{rs}) p_{\omega(r,s)}\;,
\end{align*}
where $\gamma_{rs}=\{0,1\}$, and where both $\pi(r,s)$ and $\omega(r,s)$ are permutations of the indexes $(r,s)$.
While the symmetries of $\mathcal{L}(\bm{\alpha},\bm{P})$ are automatically symmetries of the parameters, the converse is not true.
We therefore look for transformations $T$ that satisfy 
\begin{equation}
    \avg{\mathcal{L}(\bm{\alpha},\bm{P})} = \avg{\mathcal{L}\bigl(f_1(\bm{\alpha}),f_2(\bm{P})\bigr)}\;.
\end{equation}
It turns out that this constraint is satisfied if and only if $\pi=\omega$ and $\gamma_{rs}=\gamma$ $\forall (r,s)$, i.e., for transformations of the form
\begin{subequations}
\label{eq:symmetry}
\begin{align} 
    \alpha_{rs}&\mapsto \alpha_{rs}' = \alpha_{\pi(r,s)}\;,\\
    p_{rs}&\mapsto p_{rs}' = \gamma + (1 - 2\gamma) p_{\pi(r,s)}\;,
\end{align}
\end{subequations}
with $\gamma=\{0,1\}$ (see Appendix~\ref{appendix:proof_symmetries} for a detailed proof).
The permutation component of the symmetry is not to be confused with the symmetries generated by relabeling blocks: The latter only leads to $q!$ different symmetries, whereas the former correctly generates $q^*!\gg q!$ symmetries, or a total of $2q^*!$ symmetries once they are compounded with $p_{rs}\mapsto 1 -p_{rs}$.
The symmetries come about because the ordering of summation of the terms $\alpha_{rs}h(p_{rs})$ in Eq.~\eqref{eq:avg_log_likelihood_entropy_difference} does not matter, and both $h(\rho)$ and $h(p_{rs})$ are preserved when $p_{rs}\mapsto1-p_{rs}$.

As an example of symmetry, let us focus on the special transformation $p_{rs}\mapsto 1- p_{rs}$ $\forall (r,s)$ with $\pi(r,s)=(r,s)$, i.e., the only transformation that does not change the block structure of the model.
Since networks generated by these parameters can be seen as complement of one another (i.e., an edge present in $G$ is absent from $G'$, and vice-versa), we may call this transformation the \emph{graph complement} transformation.
The fact that it preserves detectability can be understood on a more intuitive level with the following argument.
Suppose that we are given an unlabeled network $G$ generated by the SBM and that we are asked to confirm or infirm the hypothesis that it was, in fact, generated by the SBM.
Even if nothing is known about the generative process, we can take the complement of the network---a trivial (and reversible) transformation.
But this should not help our cause.
After all, this transformation cannot enhance or worsen detectability since no information is added to or removed from $G$ in the process.
So we expect that $\lambda$ be preserved, and it is.
Because all other symmetries affect the block structure through a change of $\bm{\alpha}$, what the above result shows is that there is no other ``information-preserving'' transformation that can be applied to $G$ without a prior knowledge of its planted partition.

%~~~~~~~~~~~~~~~~~~~~~~~~~~~~~~~~~~~~~~~~~~~~~~~~~~~~~~~~~~~~~~~~~~~~~~~~~~~~~~
\subsubsection{Hypersurfaces and detectability regions}
\label{subsubsection:hypersurfaces}
%~~~~~~~~~~~~~~~~~~~~~~~~~~~~~~~~~~~~~~~~~~~~~~~~~~~~~~~~~~~~~~~~~~~~~~~~~~~~~~

We now turn to the problem of finding compact and explicit formulas that describe the hypersurfaces of constant $\avg{\mathcal{L}}$ in the parameter space [see Eq.~\eqref{eq:hypersurface_equation}].
In so doing we will have to be mindful of the fact that the scale $m^{\max}$ intervenes in the calculation, even though it is absent from our expression for $\avg{\mathcal{L}}$.
This can be made explicit by rewriting Eq.~\eqref{eq:hypersurface_equation} as $\avg{\log\Lambda}/m^{\max} = \widetilde{\lambda}$; it is easy to see that any given hypersurface will be comparatively closer to the region $\avg{\mathcal{L}}=0$ in larger networks.
We focus on the universal behavior of the hypersurfaces and remove all references to the scale of the problem by defining  $\lambda:=m^{\max} \widetilde{\lambda}$---predictions for real systems can be recovered by reverting to the correct scale.

While Eq.~\eqref{eq:hypersurface_equation} is easily stated, it is not easily solved for, say, $\{p_{rs}\}$.
The average normalized log-likelihood ratio involves a sum of logarithmic terms; the hypersurface equation is thus transcendental.
To further complicate matters, there are $2q^*-1=q(q-1) - 1$ degrees of freedom and the number of free parameters grow quadratically with $q$.
As a result, little can be said of truly general instances of the SBM---at least analytically.
All is not hopeless, however, because there are approximation methods that work well when the number of free parameters is not too large.
We sketch the idea here and apply it to a simpler variant of the SBM in the case study of Sec.~\ref{section:case_study}.

Expanding the binary entropy functions $h(p_{rs})$ around $p_{rs}=\rho$  $\forall r\leq s$ drastically simplifies the hypersurface equation.
Leaving the term $h(\rho)$ untouched, we find from Eq.~\eqref{eq:hypersurface_equation}
\begin{equation*}
    \lambda = h(\rho) - \sum_{r\leq s} \alpha_{rs}\left[h(\rho) + \sum_{k=1}^\infty \frac{1}{k!}\Pfn{h(x)}{x}{k} \bigg|_{x=\rho}\!\!\!\! (p_{rs}-\rho)^k\right]\;.
\end{equation*}
Due to the normalization of $\{\alpha_{rs}\}_{r\leq s}$, all terms in $h(\rho)$ cancel out, and the definition $\sum_{r\leq s}\alpha_{rs}p_{rs}=\rho$ allows us to eliminate the first order terms as well.
We are therefore left with
\begin{equation}
    \label{eq:hypersurface_simplified_equation}
    2\lambda\rho(1-\rho) = \sum_{r\leq s}\alpha_{rs}(p_{rs}-\rho)^2 + \mathcal{O}[(p_{rs}-\rho)^3]\;,
\end{equation}
where $\rho$ is fixed and $(\bm{\alpha},\bm{P})$ take on values constrained by both Eqs.~\eqref{eq:density_def} and \eqref{eq:hypersurface_simplified_equation}.
We then resort to a change of parameters and choose $\rho(\bm{P},\bm{\alpha})$ as one of the parameters.
Selecting the $q^*-1$ other parameters $\Delta_{rs}$ such that $p_{rs} = \rho(\bm{P},\bm{\alpha}) + \Delta_{rs}(\bm{P},\bm{\alpha})$, we obtain the form
\begin{equation}
    \label{eq:hypersurface_ellipsoid_equation}
    2\lambda\rho(1-\rho) = \sum_{r\leq s}\alpha_{rs}(\Delta_{rs})^2\;.
\end{equation}
Hypersurfaces are therefore ellipsoids when $p_{rs}\approx \rho\ \forall(r,s)$.

Besides the simplicity of Eq.~\eqref{eq:hypersurface_ellipsoid_equation}, there are two additional arguments for dropping the higher order terms in Eq.~\eqref{eq:hypersurface_simplified_equation}.
One, the series is invariant under the symmetry $p_{rs}\mapsto1 - p_{rs}$   $\forall (r,s)$ only if we limit ourselves to the second-order expression: It is easily verified that
\begin{multline*}
    \Pfn{h(x)}{x}{k}\bigg|_{x=\rho}(p_{rs}-\rho)^k \\= (-1)^k (k-2)!\left[\frac{1}{(\rho - 1)^{k-1} } - \frac{1}{(\rho)^{k-1}}\right](p_{rs}-\rho)^k
\end{multline*}
is off by a sign for odd powers of $k$ under the mapping $p_{rs}\mapsto 1-p_{rs}$, which also implies $\rho\mapsto 1-\rho$.
Two, the true hypersurfaces enclose sets of parameters which are convex with respect to $\bm{P}$, and so does the hypersurface  implicitly defined in Eq.~\eqref{eq:hypersurface_simplified_equation}.
The convexity of the hypersurface follows from the fact that the sublevel set of a convex function encloses a convex set \cite{Boyd2004}, and from the observation that $\avg{\mathcal{L}}$ is convex with respect to $\bm{P}$ [this is easy to show with Eq.~\eqref{eq:avg_log_likelihood_entropy_difference} and the log-sum inequality; see Appendix~\ref{appendix:convexity_proof}].
The convexity of the approximate level set is trivial to the second order, since it is an ellipsoid [Eq.~\eqref{eq:hypersurface_ellipsoid_equation}].
However, the approximate level set need not be convex when higher order terms are included.
Together, these two observations tell us that while not exact, Eq.~\eqref{eq:hypersurface_simplified_equation} captures the important \emph{qualitative} features of the problem and that it is not necessarily true of approximate solutions with only a few extra terms.

%~~~~~~~~~~~~~~~~~~~~~~~~~~~~~~~~~~~~~~~~~~~~~~~~~~~~~~~~~~~~~~~~~~~~~~~~~~~~~~
\section{Detectability distribution}
\label{section:eta-detectability}
%~~~~~~~~~~~~~~~~~~~~~~~~~~~~~~~~~~~~~~~~~~~~~~~~~~~~~~~~~~~~~~~~~~~~~~~~~~~~~~
In the previous section, we have computed the average $\avg{\mathcal{L}}$ and used it to obtain equivalence among the parameters, with respect to detectability.
We have also shown that $\avg{\mathcal{L}}>0$ for most parameters, i.e., that we could not use the necessary condition $\mathcal{L}>0$ to conclude on the \emph{undetectability} of the finite SBM, on average.
As we will now argue, this conclusion must be further refined; the full distribution of $\mathcal{L}$ leads to a more accurate picture of detectability.

%~~~~~~~~~~~~~~~~~~~~~~~~~~~~~~~~~~~~~~~~~~~~~~~~~~~~~~~~~~~~~~~~~~~~~~~~~~~~~~
\subsection{The whole picture: $\eta$--detectability}
%~~~~~~~~~~~~~~~~~~~~~~~~~~~~~~~~~~~~~~~~~~~~~~~~~~~~~~~~~~~~~~~~~~~~~~~~~~~~~~
Consider a parametrization $(\mathcal{B},\rho\bm{11}^\T+\bm{\epsilon})$ of the SBM that yields $\avg{\mathcal{L}}\approx 0$.
Turning to the distribution of $\mathcal{L}$ for this parametrization, one expects to find $\mathcal{L}<0$ with non-zero probability (unless the distribution of $\mathcal{L}$ concentrates on $\mathcal{L}=0$).
Therefore, $\avg{\mathcal{L}}$ could be indicative of detectability for some \emph{fraction} of the networks generated by the SBM.

Let us formalize this notion and introduce the concept of $\eta$--detectability.
We will say that the ensemble of networks generated with the SBM of parameters $(\mathcal{B},\bm{P})$ is $\eta$--detectable if
\begin{equation}
    \Pr(\mathcal{L}<0;\mathcal{B},\bm{P})=1 - \eta\;.
\end{equation}
That is, $\eta$ gives the fraction of networks in the ensemble which evades the necessary condition for undetectability.
If $\eta\to0$, then detection is impossible, in the sense that most instances are best described by the null hypothesis $\mathbb{Q}$.
If $\eta\to1$, then most instances contain statistical evidence for $\mathcal{B}$; detection cannot be ruled out on the basis of the log-likelihood test.

We must compute the complete distribution or the cumulative distribution function of $\mathcal{L}$ to determine $\eta$.
An exact result is out of reach since the outcome of $\mathcal{L}$ is determined by a weighted sum of independent binomial variables with non-identical distributions.
In the following paragraphs, we give an approximate derivation based on the central limit theorem---it agrees extremely well with empirical results for all but the smallest networks.

%~~~~~~~~~~~~~~~~~~~~~~~~~~~~~~~~~~~~~~~~~~~~~~~~~~~~~~~~~~~~~~~~~~~~~~~~~~~~~~
\subsection{Approximate equation for $\eta$}
%~~~~~~~~~~~~~~~~~~~~~~~~~~~~~~~~~~~~~~~~~~~~~~~~~~~~~~~~~~~~~~~~~~~~~~~~~~~~~~
\label{subsec:eta_CLT}
Equation \eqref{eq:normalized_log_likelihood} gives the normalized log-likelihood ratio as a sum of independent binomial random variables; it can be written as
\begin{subequations}
\begin{equation}
    \mathcal{L} = \sum_{r\leq s} \frac{m_{rs}}{m^{\max}}x_{rs} + C
\end{equation}
where the constants $x_{rs}$ and $C$ are given by 
\begin{align}
    x_{rs}=\log\left[\frac{p_{rs}}{\rho}\frac{1-\rho}{1-p_{rs}}\right]\;,
    \label{eq:normalized_log_likelihood_x_rs}\\
    C = \sum_{r\leq s}\alpha_{rs}\log\left[\frac{1-p_{rs}}{1-\rho}\right]\;,
    \label{eq:C_CLT}    
\end{align}
and where  $m_{rs}\sim\mathrm{Bin}(p_{rs},m^{\max}_{rs})$.

The central limit theorem (CLT) predicts that the distribution of an appropriately rescaled and centered transformation of $\mathcal{L}$ will converge to the normal distribution $\mathcal{N}(0,1)$ if the number of summed random variables $q^*=q(q+1)/2$ goes to infinity.
In the finite case, $q^*$ obviously violates the conditions of the CLT, but it  nonetheless offers a good approximation of the distribution of $\mathcal{L}$ (see Fig.~\ref{fig:CLT_quality}).

\begin{figure} 
    \centering 
    \includegraphics[width=0.95\linewidth]{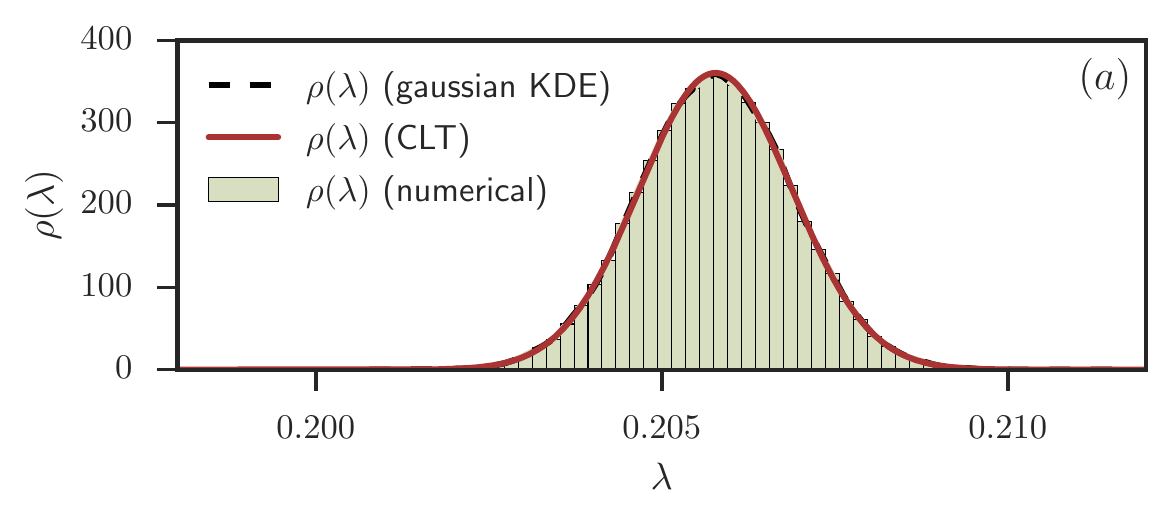}\\
    \includegraphics[width=0.95\linewidth]{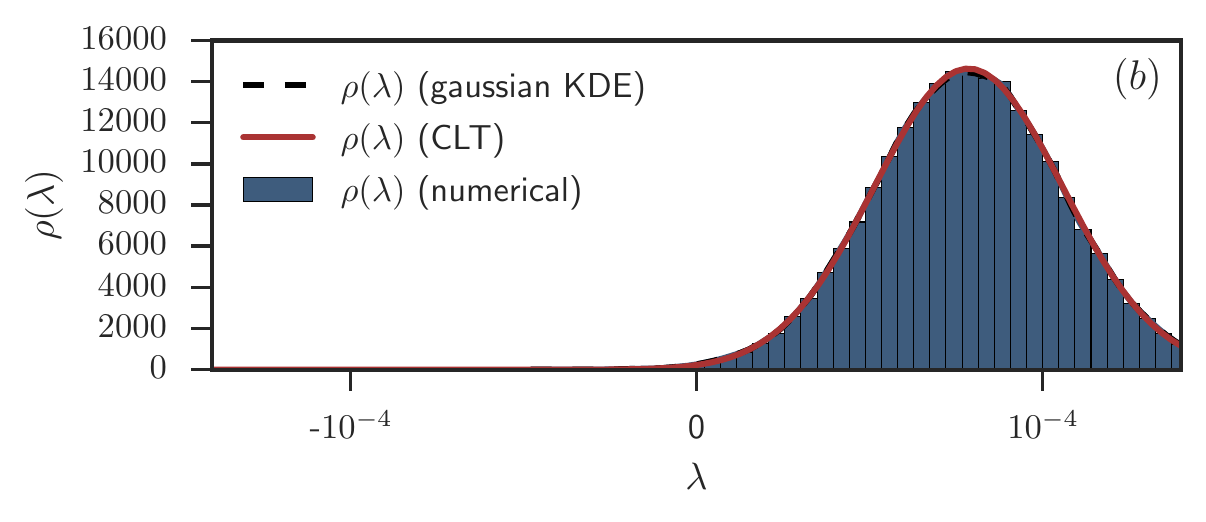} 
    \caption{Accuracy of the CLT approximation for the (a) non uniform and (b) nearly uniform SBM of Fig.~\ref{fig:log_likelihood_full_distribution}.
    Both histograms aggregate $100\;000$ samples of $\mathcal{L}$.
    The prediction of the CLT is shown in red  [see Eqs.~\eqref{eq:normalized_log_likelihood_x_rs}--\eqref{eq:recentering_CLT}].
    We plot for comparison the Gaussian kernel density estimate (KDE) of $\rho(\lambda)$ (dashed black line, hidden by the CLT curve).
    Equation~\eqref{eq:eta_approx} predicts $\eta_{(a)}=1$ and $\eta_{(b)} = 0.981(2)$; for the sample shown, numerical estimates yield  $\hat{\eta}_{(a)}=1$ and $\hat{\eta}_{(b)} = 0.980(7)$.
    }
    \label{fig:CLT_quality} 
\end{figure}

To apply the CLT, we first define the centered and normalized variable $Z=(\mathcal{L}-C - \mu_{q^*})/S_{q^*}$, where
\begin{align}
    \label{eq:var_CLT}    
    S_{q^*}^2 &= \sum_{r\leq s}\left[ \bigg\langle \left(\frac{x_{rs}m_{rs}}{m^{\max}}\right)^2\bigg\rangle\ -\ \bigg\langle \left(\frac{x_{rs}m_{rs}}{m^{\max}}\right)\bigg\rangle^2 \right]\notag\\
    &= \sum_{r\leq s} \frac{\alpha_{rs}}{m^{\max}}\ p_{rs}(1-p_{rs})x_{rs}^2
\end{align}
is the sum of the variances of the $q^*$ scaled binomial variables $x_{rs}m_{rs}/m^{\max}_{rs}$, and where
\begin{multline}
    \label{eq:recentering_CLT}
    \mu_{q^*} = \sum_{r\leq s}\mean{\frac{x_{rs}}{m^{\max}} m_{rs}} = \sum_{r\leq s} \alpha_{rs}p_{rs}x_{rs} \\ \equiv h(\rho)-\sum_{r\leq s} \alpha_{rs}h(p_{rs}) - C
\end{multline}
\end{subequations}
is the sum of their means [we have used Eq.~\eqref{eq:avg_log_likelihood_entropy_difference} in the last step].
The CLT then tells us that $Z \sim \mathcal{N}(0,1)$, approximately.

Recall that the cumulative distribution function of a normal random variable can be expressed in terms of the error function as
\begin{align}
    \Pr(Z<z) = \frac{1}{2}\left[1  + \mathrm{erf}\left(\frac{z}{\sqrt{2}}\right)\right]\;.
\end{align}
Now, assuming that $Z$ is indeed normally distributed we can use the fact that $\Pr(\mathcal{L}< 0)$ is equivalent to $\Pr[Z< -(C + \mu_{q^*})/S_{q^*}]$ to compute $\eta$.
Writing $\mu_{q^*}+C$ as $\avg{\mathcal{L}}$ [see Eq.~\eqref{eq:recentering_CLT}], we find
\begin{equation}
    \label{eq:eta_approx}
    \eta \approx \frac{1}{2}\left[1 + \mathrm{erf}\left(\frac{\avg{\mathcal{L}}}{\sqrt{2}S_{q^*}}\right)\right]\;,
\end{equation}
i.e., an (approximate) equation in closed form for $\eta$.

Crucially, Eq.~\eqref{eq:eta_approx} predicts that $\eta$ can never be smaller than $1/2$.
This comes about because (i) $\avg{\mathcal{L}}>0$ and (ii) $S_{q^*}$ is a sum of variances, i.e., a positive quantity.
There are therefore two possible limits which will yield $\avg{\mathcal{L}}/S_{q^*}\approx 0$ and $\eta=1/2$: Either $\avg{\mathcal{L}}=0$ or $S_{q^*}\gg 0$.
Some care must be exerted in analyzing the case $\avg{\mathcal{L}}=0$;
Eqs.~\eqref{eq:normalized_log_likelihood} and \eqref{eq:KL_divergence} tell us that the distribution of $\mathcal{L}$ is concentrated on $0$ when its average is exactly equal to 0.
We conclude that $\eta=1/2$ is never reached but only approached asymptotically, for parameters that yield $\avg{\mathcal{L}} = \varepsilon$, with $\varepsilon$ small but different from zero.
The consequence of $\eta\geq 1/2$ is that at most half of the instances of the SBM can be declared undetectable on the account of the condition $\mathcal{L}<0$.

%~~~~~~~~~~~~~~~~~~~~~~~~~~~~~~~~~~~~~~~~~~~~~~~~~~~~~~~~~~~~~~~~~~~~~~~~~~~~~~
\subsection{Relation between average detectability and $\eta$--detectability}
%~~~~~~~~~~~~~~~~~~~~~~~~~~~~~~~~~~~~~~~~~~~~~~~~~~~~~~~~~~~~~~~~~~~~~~~~~~~~~~

We can immediately reach a few conclusions on the interplay between the notions of average and $\eta$--detectability.
First, the symmetries of $\avg{\mathcal{L}}$, (see Sec.~\ref{subsubsection:symmetries_avg_detectability}) translates into symmetries for $\eta$.
To see this, first notice that $S_{q^*}$ is conserved under the mapping $p_{rs}\mapsto 1 - p_{rs}$
\begin{align*}
  [x_{rs}(p_{rs},\rho)]^2 &\mapsto [-x_{rs}(1-p_{rs},1-\rho)]^2\;,\\
  p_{rs}(1-p_{rs}) &\mapsto (1-p_{rs})p_{rs}\;.
\end{align*}
and that a permutation of the indexes $\pi(r,s)$ only changes the order of summation of the terms of $S_{q^*}$.
Second, hypersurfaces of constant average detectability need not be hypersurfaces of constant $\eta$--detectability.

To investigate this second important aspect of the connection between average detectability and $\eta$--detectability, let us further approximate Eq.~\eqref{eq:eta_approx}.
The MacLaurin series of the error function is, to the first order,
\begin{align}
    \eta &= \frac{1}{2}\left\{1 + \frac{2}{\sqrt{\pi}}\left[\frac{\avg{\mathcal{L}}}{S_{q^*}} - \mathcal{O}\bigl(\avg{\mathcal{L}}^3/S_{q^*}^3\bigr)\right]\right\}\notag\;,\\
    &\approx \frac{1}{\sqrt{2\pi}} \frac{\avg{\mathcal{L}}}{S_{q^*}} + \frac{1}{2}\;.
    \label{eq:eta_maclaurin}
\end{align}
This is a reasonably accurate calculation of $\eta$ when $\avg{\mathcal{L}}$ is small, i.e., close to the \emph{average} undetectable regime. (Recall that we do not allow diverging $S_{q^*}$ for the reasons stated in Sec.~\ref{subsec:eta_CLT}).
It then becomes clear that on the hypersurfaces where $\avg{\mathcal{L}}=\lambda$ is constant (and close to 0), 
\begin{equation}
    \label{eq:nearly_random_variance_tradeoff}
    \sqrt{2\pi}\left(\eta-\frac{1}{2}\right)S_{q^*} = \lambda\;,
\end{equation} 
is conserved rather than $\eta$ itself.
Equation \eqref{eq:nearly_random_variance_tradeoff} embodies a trade-off between accuracy ($\eta$) and variance ($S_{q^*}$): In the regions of the hypersurface of constant $\avg{\mathcal{L}}$ where the variance is large, $\eta$ must be comparatively small, and vice-versa.

%~~~~~~~~~~~~~~~~~~~~~~~~~~~~~~~~~~~~~~~~~~~~~~~~~~~~~~~~~~~~~~~~~~~~~~~~~~~~~~
\subsection{1--detectability}
%~~~~~~~~~~~~~~~~~~~~~~~~~~~~~~~~~~~~~~~~~~~~~~~~~~~~~~~~~~~~~~~~~~~~~~~~~~~~~~
Now, turning to the complementary case where $\avg{\mathcal{L}}$---and consequently $\eta$---is close to its maximum, we obtain a simple criterion for 1--detectability based the asymptotic behavior of $\mathrm{erf}(x)$.
It is reasonable to define a (small) threshold $T$ beyond which $\mathrm{erf}(x>T) = 1$ for all practical purposes.
The error function goes asymptotically to $1$ with large values of its argument, but reaches its maximum of $\mathrm{erf}(x)=1$ very quickly, so quickly, in fact, that $\mathrm{erf}(5)$ is numerically equal to 1 to the 10\textsuperscript{th} decimal place.

Asking that the argument of $\mathrm{erf}(x)$ in Eq.~\eqref{eq:eta_approx} be greater than this practical threshold, we obtain the inequality
\begin{equation}
    \label{eq:approximate_1_detectable}
    \avg{\mathcal{L}} \geq \sqrt{2}T S_{q^*}
\end{equation}
for $1$--detectability.
Whenever the inequality holds, the associated ensemble is 1--detectable with a tolerance threshold $T$, i.e., we can say that for all practical purposes, there are no instances of the SBM that are necessarily \footnote{Since $\mathcal{L}>0$ is not sufficient for detectability, some instances could still be undetectable.} undetectable.

%~~~~~~~~~~~~~~~~~~~~~~~~~~~~~~~~~~~~~~~~~~~~~~~~~~~~~~~~~~~~~~~~~~~~~~~~~~~~~~
\section{Case study: General Modular Graphs}
\label{section:case_study}
%~~~~~~~~~~~~~~~~~~~~~~~~~~~~~~~~~~~~~~~~~~~~~~~~~~~~~~~~~~~~~~~~~~~~~~~~~~~~~~
The stochastic block model encompasses quite a few well-known models as special cases;
noteworthy examples include the \emph{planted partition model} \cite{Jerrum1998,Condon2001}, the closely related \emph{symmetric SBM} (SSBM) \cite{Abbe2016,Nadakuditi2012,Decelle2011a}, the \emph{core-periphery model}  \cite{Borgatti2000}, and many more.
These simplified models are important for two reasons.
One, they are good abstractions of structural patterns found in real networks, and a complete understanding of their behavior with respect to detectability is therefore crucial.
Two, they are simple enough to lend themselves to a thorough analysis; this contrasts with the general case, where simple analytical expressions are hard to come by.

In the paragraphs that follow,  we investigate the \emph{general modular graph model} (GMGM) \cite{Kawamoto2017}, a mathematically simple, yet phenomenologically rich simplified model.
Thanks to its simpler parametrization, we obtain easily interpretable versions of the expressions and results derived in  Secs.~\ref{section:finite_limit}--\ref{section:eta-detectability}.

\begin{figure*}
    \includegraphics[height=11\baselineskip,valign=t]{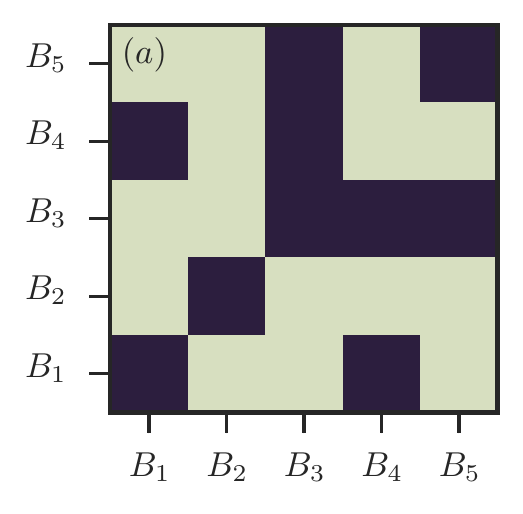}
    \includegraphics[height=12\baselineskip,valign=t]{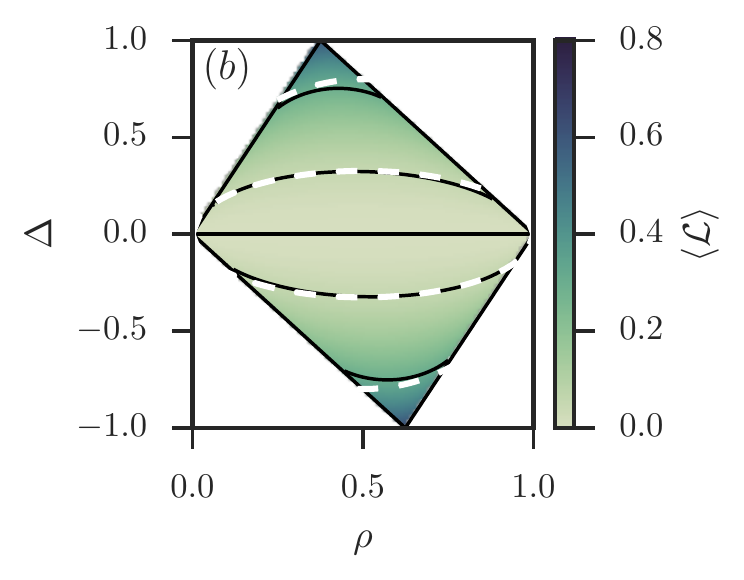}
    \includegraphics[height=12\baselineskip,valign=t]{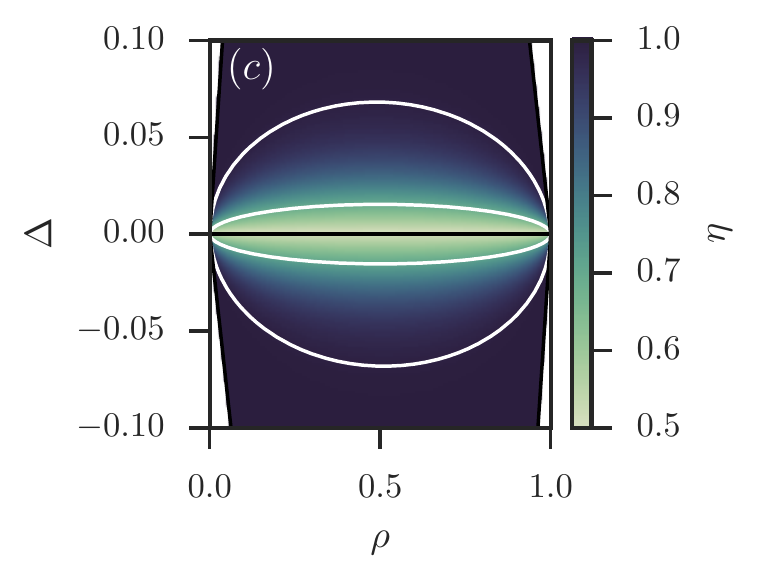}
    \caption{(color online)
    Detectability in the general modular graph model.
    All figures use the same indicator matrix $\bm{W}$ [panel (a)] and the size vector $\bm{n} = [10,30,20,20,20]$ ($n=100$ nodes).
    (a) Example of density matrix $\bm{P}$ allowed in the GMGM. Dark squares indicate block pairs of the inner type and light squares indicate pairs of the outer type.
    (b) Average detectability in the density space of the GMGM.
    Both the numerical solution of $\avg{\mathcal{L}}=\lambda$ (solid black line) and the prediction of Eq.~\eqref{eq:explicit_hypersurface_GMG} (dashed white line) are shown, for $\lambda=0.05$ and $0.3$.
    (c) $\eta(\rho,\Delta;\beta)$ in the density space of the GMGM; notice the change of $\Delta$--axis.
    Solid white lines are curves where $\eta(\rho,\Delta;\beta)=\eta^*$, with $\eta^{*}=0.7$ (central curve) and $\eta^{*}=0.99$  (outer curve).
    Equation~\eqref{eq:eta_approx} is used to compute both $\eta$ and $\eta^*$.
    }
    \label{fig:general_modular_graph}
\end{figure*}

%~~~~~~~~~~~~~~~~~~~~~~~~~~~~~~~~~~~~~~~~~~~~~~~~~~~~~~~~~~~~~~~~~~~~~~~~~~~~~~
\subsection{Parametrization of general modular graphs}
%~~~~~~~~~~~~~~~~~~~~~~~~~~~~~~~~~~~~~~~~~~~~~~~~~~~~~~~~~~~~~~~~~~~~~~~~~~~~~~
The GMGM can be seen as constrained version of the SBM, in which \emph{pairs} of blocks assume one of two roles: Inner or outer pairs.
If a pair of blocks $(B_r, B_s)$ is of the ``inner type'', then one sets $p_{rs}=p_{\ins}$.
If a pair of blocks $(B_r, B_s)$ is of the ``outer type'', then one sets $p_{rs}=p_{\outs}$.
The resulting density matrices can therefore be expressed as
\begin{equation}
    \label{eq:prob_matrix_gmg}
    \bm{P} = (p_{\ins}-p_{\outs})\bm{W} + p_{\outs}\bm{11}^\T\;,
\end{equation}
where  $\bm{W}$ is a $q\times q$ indicator matrix [$w_{rs}=1$ if $(B_r,B_s)$ is an inner pair], and where $\bm{1}$ is a length $q$ vector of ones.
A non-trivial example of a density matrix of this form is shown in Fig.~\ref{fig:general_modular_graph} (a).
The figure is meant to illustrate just how diverse the networks generated by the GMGM may be, but it is also important to note that the results of this section apply to \emph{any} ensemble whose density matrix can be written as in Eq.~\eqref{eq:prob_matrix_gmg}.
This includes, for example, the $q$--block SSBM, a special case of the GMGM obtained by setting $\bm{W}=\bm{I}_q$ and $\{n_r=n/q\}_{r=1,..,q}$ (see Ref.~\cite{Abbe2016b} for a discussion of the SSBM).

While the parametrization in terms of $p_{\ins}$ and $p_{\outs}$ is simple, we will prefer an arguably more convoluted parametrization which is also more revealing of the natural symmetries of the GMGM (in line with the transformation proposed in Sec.~\ref{subsubsection:hypersurfaces}).
The first natural parameter is the average density, which can be computed from Eqs.~\eqref{eq:density_def} and \eqref{eq:prob_matrix_gmg} and which equals
\begin{subequations}
\label{eq:natural_gmg_params}
\begin{align}
    \label{eq:density_GMG}
    \rho&=\sum_{r\leq s} \alpha_{rs}[w_{rs}p_{\ins} + (1-w_{rs})p_{\outs}]\;,\notag\\
        &=\beta p_{\ins} + (1-\beta)p_{\outs}\;,
\end{align}
where $\beta:=\sum_{r\leq s} \alpha_{rs} w_{rs}$ is the fraction of \emph{potential} edges that falls between block pairs of the inner type.
The second natural parameter is simply the difference
\begin{equation}
    \Delta =p_{\ins} - p_{\outs}\;.
    \label{eq:delta_GMG}
\end{equation}
\end{subequations}
The absolute value of $\Delta$ quantifies the distance between the parameters of the GMGM and that of the equivalent random ensemble;
its sign tells us which type of pairs is more densely connected.
In this natural parametrization, the density matrix takes on the form $\bm{P} = \rho\bm{11}^\T +\Delta(1-\beta)\bm{W}$, i.e., a uniform matrix of $\rho$ with perturbation proportional to $\Delta(1-\beta)$ for the inner pairs.
It might appear that we have increased the complexity of the model description, since the additional parameter $\beta$ now appears in the definition of the density matrix.
It is, however, not the case, because we could consider the combined parameter $\widetilde{\Delta}=\Delta(1-\beta)$.
Therefore, Eqs.~\eqref{eq:density_GMG} and \eqref{eq:delta_GMG}, together with $\bm{W}$ and $\bm{n}$, suffice to unambiguously parametrize the model.

%~~~~~~~~~~~~~~~~~~~~~~~~~~~~~~~~~~~~~~~~~~~~~~~~~~~~~~~~~~~~~~~~~~~~~~~~~~~~~~
\subsection{Average detectability of general modular graphs}
%~~~~~~~~~~~~~~~~~~~~~~~~~~~~~~~~~~~~~~~~~~~~~~~~~~~~~~~~~~~~~~~~~~~~~~~~~~~~~~
The average normalized log-likelihood ratio $\avg{\mathcal{L}}$ is tremendously simplified in the natural parametrization of the GMGM;
it is straightforward to show that the ratio takes on the compact (and symmetric) form
\begin{multline}
    \label{eq:gmg_likelihoodratio_per_edges}
    \avg{\mathcal{L}(\rho,\Delta;\beta)} = \beta\Big\{h(\rho) -  h\bigl[\rho+(1-\beta)\Delta\bigr]\Big\} \\+ (1-\beta)\Big\{h(\rho) -  h\bigl[\rho-\beta\Delta\bigr]\Big\} \;,
\end{multline}
by using $p_{rs} = w_{rs}p_{\ins} + (1-w_{rs})p_{\outs}$ together with the inverse of Eqs.~\eqref{eq:density_GMG} and \eqref{eq:delta_GMG},
\begin{subequations}
\label{eq:natural_gmg_params_inverse}
\begin{align}
    p_{\ins} &= \rho + (1-\beta)\Delta\;,\\
    p_{\outs} &= \rho -\beta\Delta\;.
\end{align}
\end{subequations}

In Fig.~\ref{fig:general_modular_graph} (b), we plot $\avg{\mathcal{L}(\rho,\Delta;\beta)}$ in the $(\rho,\Delta)$ space---hereafter the density space---for the indicator matrix $\bm{W}$ shown in Fig.~\ref{fig:general_modular_graph} (a) (and unequal block sizes, see caption).
Unsurprisingly, $\avg{\mathcal{L}}$ is largest when the block types are clearly separated from one another, i.e., when $|\Delta|$ is the largest.
Notice, however, how large separations are \emph{not} achievable for dense or sparse networks.
This is due to the fact that not all $(\rho,\Delta)$ pairs map to probabilities $(p_{\ins},p_{\outs})$ in $[0,1]$.
The region of the density space that \emph{does} yield probabilities is the interior of the quadrilateral whose vertices are, in $(\rho,\Delta)$ coordinates: $(0,0), (\beta,1), (1,0), (1-\beta,-1)$. 
Changing the value of $\beta$ skews this accessible region and, presumably, the functions that are defined on it, such as $\avg{\mathcal{L}(\rho,\Delta;\beta)}$.

We also show on Fig.~\ref{fig:general_modular_graph} (b) two members of the level set defined by $\avg{\mathcal{L}(\rho,\Delta;\beta)}=\lambda$.
As mentioned previously, the exact functional form of this family of hypersurfaces (here simply curves) seems elusive, but an approximate solution is available.
Using the method highlighted in Sec.~\ref{section:average_detectability}, we find, to the second order,
\begin{align}
    2\lambda\rho(1-\rho) &\approx \sum_{r\leq s}\alpha_{rs}(p_{rs}-\rho)^2 \notag\\
                         &= \beta[(1-\beta)\Delta]^2 + (1-\beta)(\beta\Delta)^2\;.
    \label{eq:hypersurface_GMG}
\end{align}
Equation \eqref{eq:hypersurface_GMG} fixes the relative value of all parameters on the line where $\avg{\mathcal{L}} = \lambda$.
Solving for $\Delta$, we find
\begin{equation}
    \Delta^*(\rho;\lambda,\beta) = \pm \sqrt{2 \lambda \frac{\rho(1-\rho)}{\beta(1-\beta)}}\;,
    \label{eq:explicit_hypersurface_GMG}
\end{equation}
also shown on Fig.~\ref{fig:general_modular_graph} (b) for comparison.

Figure \ref{fig:general_modular_graph} highlights the accuracy of our approximation when $\lambda$ is small.
But it also highlights its inaccuracy when $\lambda$ is large; $\lambda \gg 1$ forces $\Delta^*(\rho;\lambda,\beta)$ to pass through a region where $\Delta^*  \approx 1$, i.e., a region where the omitted terms on the right-hand-side of Eq.~\eqref{eq:hypersurface_GMG} contribute heavily.
Fortunately, this is not so problematic, since most detectability related phenomena---phase transitions, undetectable instances, etc.---arise near $\Delta=0$, i.e., where the approximation works.

%~~~~~~~~~~~~~~~~~~~~~~~~~~~~~~~~~~~~~~~~~~~~~~~~~~~~~~~~~~~~~~~~~~~~~~~~~~~~~~
\subsection{$\eta$--detectability  of general modular graphs}
%~~~~~~~~~~~~~~~~~~~~~~~~~~~~~~~~~~~~~~~~~~~~~~~~~~~~~~~~~~~~~~~~~~~~~~~~~~~~~~
While $\avg{\mathcal{L}(\rho,\Delta;\beta)}$ takes on a particularly compact form once we substitute $\{p_{rs}\}$ by the natural parameters of the GMGM, the same cannot be said of $\eta(\rho,\Delta;\beta,n)$.
Some analytical progress can be made by, e.g., noticing that only two types of terms are involved in the calculation of $S_{q^*}$, but, ultimately, the resulting expression is no more useful than the simple Eqs.~\eqref{eq:eta_approx} and \eqref{eq:eta_maclaurin}.
We will, therefore, omit the calculation of $\eta$.

In Fig.~\ref{fig:general_modular_graph}~(c) we plot $\eta(\rho,\Delta;\beta,n)$ in the density space [using Eq.~\eqref{eq:eta_approx}].
We also display the numerical solutions of $\eta(\rho,\Delta;\beta,n)=\eta^*$ for two values of $\eta^*$.
The figure highlights just how quickly $\eta$ goes to 1 as a function of $\Delta$, even for the fairly small system sizes considered:
We find that $\eta\geq 0.99$ for \emph{any} value of $\rho$, as soon as $\Delta > 0.06$.
The condition in Eq.~\eqref{eq:ratio_randomvariable} is therefore a \emph{weak} one.
It allows us to determine that some parameters are overwhelmingly undetectable, but only when $\Delta$ is very close to 0.

Figure \ref{fig:general_modular_graph} also shows how increases in variance translate into decreases in accuracy  [see Eq.~\eqref{eq:nearly_random_variance_tradeoff}]:
Following a line of constant (and relatively small) $\Delta$, one can  see that $\eta$ is minimized close to $\rho=1/2$, i.e., near the maximum of variance.
This is characteristic of many parametrizations of the SBM and GMGM; it turns out that, for fixed $n$, impossible detection problems are not confined to vanishing densities.
In fact, values of $\rho$ closer to $1/2$ are associated with a comparatively larger interval of $\Delta$ for which detection is impossible.

%~~~~~~~~~~~~~~~~~~~~~~~~~~~~~~~~~~~~~~~~~~~~~~~~~~~~~~~~~~~~~~~~~~~~~~~~~~~~~~
\subsection{Symmetries of general modular graphs}
%~~~~~~~~~~~~~~~~~~~~~~~~~~~~~~~~~~~~~~~~~~~~~~~~~~~~~~~~~~~~~~~~~~~~~~~~~~~~~~
In Secs.~\ref{section:average_detectability} and \ref{section:eta-detectability}, we have proved that there are  $2q^*!$ transformations that preserve $\avg{\mathcal{L}(\rho,\Delta;\beta)}$ and $\eta(\rho,\Delta;\beta,n)$.
We could therefore go about computing the symmetries of the GMGM by listing all of these transformations in terms of $(\rho,\Delta,\beta)$.
But since there are only three free parameters in the GMGM, we can also choose an alternative route and directly solve $\avg{\mathcal{L}(\rho,\Delta;\beta)} = \avg{\mathcal{L}(a_1\rho+b_1,a_2\Delta+b_2;a_3\beta + b_3)}$ by, e.g., obtaining a linear system from the Taylor series of $\avg{\mathcal{L}(\rho,\Delta;\beta)}$.
This simpler approach yields the following set of $\lambda$--preserving transformations for the model:
\begin{subequations}
\label{eq:gmg_syms}
\begin{align}
    (\rho,\Delta,\beta) &\mapsto (\rho,\Delta,\beta)\;,  \label{eq:gmg_syms_a}\\
    (\rho,\Delta,\beta) &\mapsto (\rho,-\Delta,1-\beta)\;,   \label{eq:gmg_syms_b}\\
    (\rho,\Delta,\beta) &\mapsto (1-\rho,\Delta,1-\beta)\;,  \label{eq:gmg_syms_c}\\
    (\rho,\Delta,\beta) &\mapsto (1-\rho,-\Delta,\beta)\;.   \label{eq:gmg_syms_d}
\end{align}
\end{subequations}
It is straightforward to check that these transformations form a group, whose product is the composition of two transformations.
A Cayley table reveals that the group is isomorphic to the Klein four-group $Z_2\times Z_2$.

One immediately notices a large gap between the number of symmetries predicted by the calculations of Sec.~\ref{subsubsection:symmetries_avg_detectability} ($2q^*!$) and the number of symmetries appearing in Eq.~\eqref{eq:gmg_syms} ($4$, independent of $q$).
The gap is explained by the fact that every symmetry of the general SBM maps onto one of the four transformations listed in Eq.~\eqref{eq:gmg_syms} \footnote{Another explanation is that there are effectively $q^*=2$ pairs of blocks in the eyes of our formalism: A single inner pair and a single outer pair, with, respectively, a fraction $\beta$ and $1-\beta$ of all possible edges.}
A sizable fraction of the symmetries reduce to Eq~\eqref{eq:gmg_syms_a}, since permutations $\pi(r,s)$ cannot modify the natural parameters of the GMGM: The type of block pair $(B_r, B_s)$---characterized by $p_{rs}$---is permuted alongside its share of potential edges $\alpha_{rs}$.
Another important fraction of the symmetries is accounted for by the ``graph complement transformation'': Any transformation $\bm{P}=\bm{11}^\T -\bm{P}$ plus a permutation reduces to Eq.~\eqref{eq:gmg_syms_d}.
This leaves two symmetries, which happen to be artifacts of our choice of nomenclature.
To see this, \emph{rename} pair types, i.e., call inner pairs ``outer pairs'' and vice-versa.
Neither the density $\rho$ nor $|\Delta|$ will change.
But both the sign of $\Delta$ and the value of $\beta$ will be altered.
With this in mind, it becomes clear that Eq.~\eqref{eq:gmg_syms_b} corresponds to the permutation symmetry, and that Eq.~\eqref{eq:gmg_syms_c} corresponds to the graph complement symmetry, both up to a renaming of the types.

%~~~~~~~~~~~~~~~~~~~~~~~~~~~~~~~~~~~~~~~~~~~~~~~~~~~~~~~~~~~~~~~~~~~~~~~~~~~~~~
\subsection{Where the framework is put to the test: Inference}
\label{subsection:inference_on_ssbm}
%~~~~~~~~~~~~~~~~~~~~~~~~~~~~~~~~~~~~~~~~~~~~~~~~~~~~~~~~~~~~~~~~~~~~~~~~~~~~~~
\subsubsection{Procedure}
It will be instructive to put our framework to the test and compare its predictions with numerical experiments that involve inference, i.e., the detection of the planted partition of actual instances of the GMGM.
We will use the following procedure: (i) generate an instance of the model, (ii) run an inference algorithm on the instance, and (iii) compute the correlation of the inferred and planted partition (see below for a precise definition).
The average detectability $\avg{\mathcal{L}}$ should bound the point where the average correlation becomes significant, and $\eta$--detectability should give an upper bound on the fraction of correlated instances.

Even for the small size considered, it is impossible to compute all quantities involved in the process exactly; we therefore resort to sub-sampling.
We use an efficient algorithm \footnote{We give a reference implementation of the algorithm in C++ at \url{www.github.com/jg-you/sbm_canonical_mcmc}.} based on the Metropolis-Hastings algorithm of Ref.~\cite{Snijders1997}, which, unlike belief propagation \cite{Decelle2011a}, works well for dense networks with many short loops.
The principle of the algorithm is to construct an ergodic chain of partitions $\mathcal{B}_0,...,\mathcal{B}_T$, and to sample from the chain to approximate the probability
\begin{equation}
  \mu_i^r(G) = \sum_{\{\mathcal{B}_\sigma\}} \Pr(\mathcal{B}_\sigma|G,\bm{P},\bm{n}) \delta(\sigma(v_i) = r) \label{eq:marginal}
\end{equation}
that node $v_i$ is in block $B_r$, given a network $G$ and some parameter $\bm{P}$ and $\bm{n}$.
It is easy to see that one can then maximize the probability of guessing the partition correctly by assigning nodes according to \cite{Decelle2011b}
\begin{equation}
  \hat{\sigma}(v_i) = \mathrm{argmax}_{r} (\mu_{i}^{r} )\;.\label{eq:optimal_guess}
\end{equation}
We choose a simple set of moves that  yields an ergodic chain over all $\{\mathcal{B}\}$: at each step, we change the block of a randomly selected node $v_i$ from $\sigma(v_i)=B_r$ to a randomly and uniformly selected block $B_s$, with probability $\min\{1,\mathcal{A}\}$, where
\begin{multline}
  \mathcal{A} = 
  \left[\frac{p_{rs}(1-p_{rr})}{p_{rr}(1-p_{rs})}\right]^{k_r^{(i)}}\!\!\left[\frac{p_{ss}(1-p_{rs})}{p_{rs}(1-p_{ss})}\right]^{k_s^{(i)}}\!\!
  \\\times\left[\frac{1-p_{rs}}{1-p_{rr}}\right]^{n_r -1}\!\!\left[\frac{1-p_{ss}}{1-p_{rs}}\right]^{n_s}
  \\\times \prod_{l\neq r, s} \left[\frac{p_{ls}(1-p_{rl})}{p_{rl}(1-p_{ls})}\right]^{k_l^{(i)}}\!\!\!\!\left[\frac{1-p_{ls}}{1-p_{rl}}\right]^{n_l}\;,
  \label{eq:a_sbm_single}
\end{multline}
and $k^{(i)}_l$ the number of neighbors of node $v_i$ in block $B_l$ \cite{Snijders1997}.
The space of all partitions is obviously connected by this move set, and the possibility of resampling a configuration ensures that the chain is aperiodic.
Furthermore, since transition probabilities are constructed according to the prescription Metropolis-Hastings, the chain is ergodic and samples from $\mathbb{P}(\mathcal{B}|G,\bm{P},\bm{n})$.
Note that we assume that $\bm{P}$ is known when we compute Eq.~\eqref{eq:marginal}.
Learning the parameters can be done separately, see Ref.~\cite{Decelle2011b}, for example.

In the spirit of Refs.~\cite{Decelle2011b,Decelle2011a}, we 
initialize the algorithm with the planted partition itself.
This ensures that we will achieve the information-theoretic threshold, even if efficient inference is impossible  \cite{Decelle2011b}.
To see this, first consider the case where the planted partition is information-theoretically detectable.
In this case, the chain will concentrate around the initial configuration, and the marginal distribution [Eq.~\eqref{eq:marginal}] will yield a distribution correlated with the planted partition.
We will have to proceed with care, however, since two scenarios may occur in the information-theoretically undetectable phase.
If there is no hard phase---e.g., when $q=2$ \cite{Mossel2013}---, the algorithm will show no particular preference for the initial configuration and wander away toward partitions uncorrelated with the planted partition.
But if there is a hard phase, one will have to wait for a period that diverges exponentially in the system size before the sampler becomes uncorrelated with its initial state \cite{Decelle2011b}.
This complicates convergence diagnosis and can lead one to conclude that correlated inference is possible even though it's not. 
To avoid these difficulties, we will simply restrict ourselves to the cases where the hard phase does not exist \cite{Abbe2016b}.

Once the estimated partition $\hat{\mathcal{B}}$ is obtained via Eq.~\eqref{eq:optimal_guess}, we compute its correlation with  $\mathcal{B}$---the planted partition---using a measure that accounts for finite-size effects.
The so-called relative normalized mutual information (rNMI) of Ref.~\cite{Zhang2015} appears a good choice.
Much like the well-known NMI \cite{Danon2005,Kvalseth1987}, the rNMI is bounded to the [0, 1] interval, and $\mathrm{rNMI}(\mathcal{B}_p,\hat{\mathcal{B}})=1$ means that the planted partition  $\mathcal{B}_p$ and the inferred partition $\hat{\mathcal{B}}$ are identical.
Unlike the NMI, $\mathrm{rNMI}(\mathcal{B}_p,\hat{\mathcal{B}})=0$ signals the absence of correlation between the two partitions, even in finite networks.

\begin{figure}[!h]
    \centering
    \includegraphics[width=0.74\linewidth]{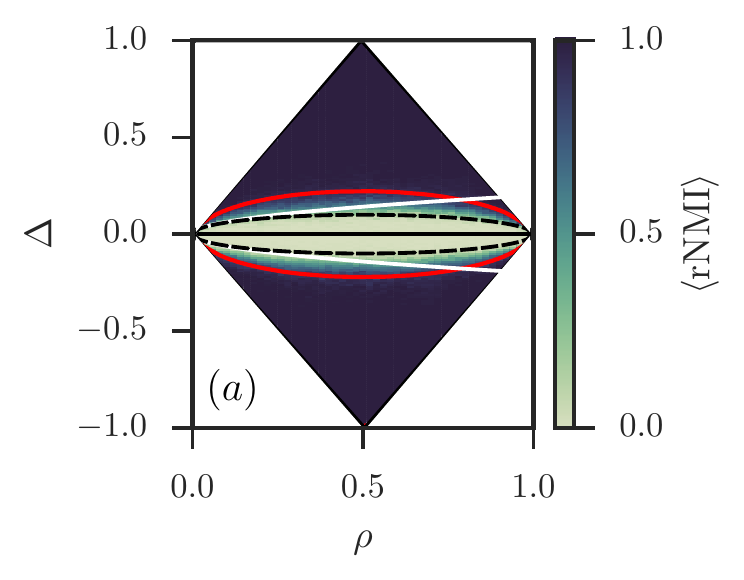}
    \includegraphics[width=0.95\linewidth]{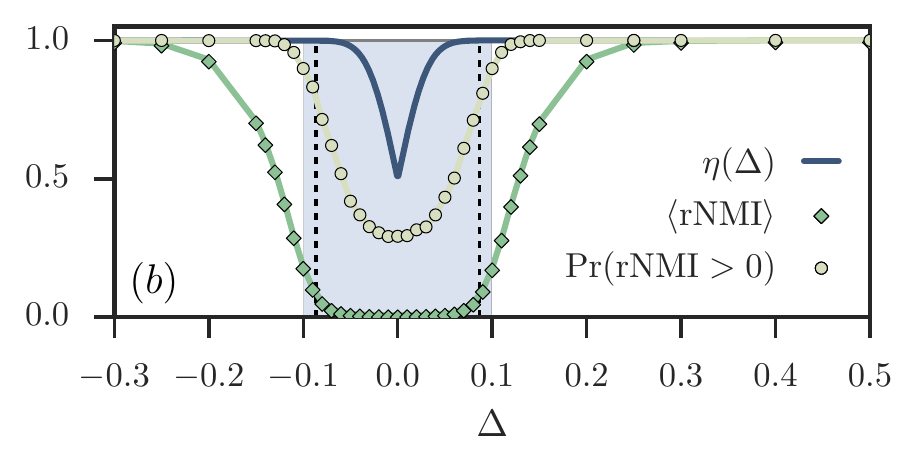}
    \includegraphics[width=0.95\linewidth]{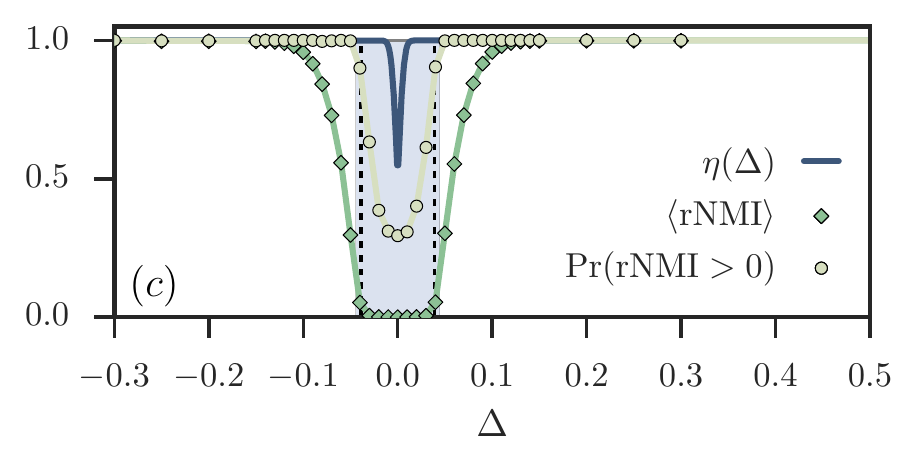}
    \caption{Inference of the GMGM.
    All figures show results for the special case $q=2$, $\bm{W}=\bm{I}_2$, and $\bm{n}=[n/2, n/2]$, corresponding to the $q=2$ SSBM \cite{Abbe2016b}.
    All empirical results are averaged over $10^4$ independent instances of the SBM.
    (a) Average rNMI of the planted and the inferred partition in the density space of the model of size $n=100$. 
    Solid red lines mark the boundaries of the 1--detectability region, with tolerance threshold $T=4\sqrt{2}$; see Eq.~\eqref{eq:approximate_1_detectable}.
    Dotted black lines show the two solutions of $\Delta^*(\rho;\lambda=1/2n,\beta)$; see Eq.~\eqref{eq:explicit_hypersurface_GMG}.
    White lines show the finite-size Kesten-Stigum (KS) bound, before it is adjusted for the symmetries of the problem.
    (b, c) Phase transition at constant $\rho=0.25$ for networks of $n=100$ nodes (b) and $n=500$ nodes (c).
    Circles indicate the fraction of instances for which a correlated partition could be identified, while diamonds show the average of the rNMI (lines are added to guide the eye).
    Blue solid curves show $\eta(\Delta;\rho,\beta,n)$; see Eq.~\eqref{eq:eta_approx}.
    The shaded region lies below the finite-size KS bound $\Delta=\pm q\sqrt{\rho/n}$ (here with $q=2$).
    The dotted lines show the two solutions of $\Delta^*(\rho;\lambda=1/2n,\beta=1/2)$.
    } 
    \label{fig:general_modular_graph_transition}
\end{figure}

\subsubsection{Results}
In Fig.~\ref{fig:general_modular_graph_transition}~(a), we plot $\avg{\mathrm{rNMI}(\mathcal{B}_p,\hat{\mathcal{B}})}$ in the density space of the GMGM.
We use the parameters $\bm{W}=\bm{I}$, and $\bm{n}=[n/2,n/2]$ (i.e., the SSBM), since the resulting ensemble is conjectured to be the hardest of all, with respect to detectability \cite{Decelle2011b}.
Two important parallels can be drawn between the results shown in Fig.~\ref{fig:general_modular_graph_transition}~(a) and the functional form of $\avg{\mathcal{L}(\rho,\Delta;\beta)}$ and $\eta(\rho,\Delta;\beta,n)$ [shown in Figs.~\ref{fig:general_modular_graph}~(b) and \ref{fig:general_modular_graph}~(c) for a different GMGM].
First, notice how the boundary that marks the onset of the (theoretically) 1--detectable region partitions the density space in two qualitative regimes: A regime where perfect detection is possible \emph{for all instances}, and a region where it is not.
There is, of course, some level of arbitrariness involved in selecting the threshold $T$ [see Eq.~\eqref{eq:approximate_1_detectable}].
But the fact that a line of constant $\eta$ partitions the space is a hint that while $\mathcal{L}<0$ is not sufficient for undetectability, there exists a level of significant $\lambda^*$ for which $\mathcal{L}$ properly separates detectable and undetectable instances.

The second important parallel concerns hypersurfaces of constant $\avg{\mathcal{L}}$ and their connection with $\avg{\mathrm{rNMI}}$.
We have argued in Sec.~\ref{section:average_detectability} that $\avg{\mathcal{L}}$ is a good predictor of the accuracy of an optimal inference algorithms (with potentially exponential  complexity).
It should, therefore, not be surprising that there is an hypersurface of constant $\avg{\mathcal{L}}$ which \emph{also} partitions the density space in two qualitative regions \footnote{We do not have a procedure to determine the value of $\lambda$ within the information-theoretical framework itself. However, random matrix theory and recent developments in Information theory offers some insights as to why one should have $\lambda\propto 1/n$, see Appendix \ref{appendix:connection_with_rmt} and Ref.~\cite{Lesieur2015}  for details.}: One where $\avg{\mathrm{rNMI}}\approx 0$ and one where $\avg{\mathrm{rNMI}}$ is clearly greater than zero.
On this hypersurface, the average level of significance is the same for all parametrizations of the GMGM; our results show that the inference algorithm achieves correspondingly uniform accuracy for all parameters on the surface.

One could argue that these parallels are not so obvious in Fig.~\ref{fig:general_modular_graph_transition}~(a);
we therefore focus on a subset of the density space in Figs.~\ref{fig:general_modular_graph_transition}~(b) and \ref{fig:general_modular_graph_transition}~(c) to make our case clearer.
In these figures, we plot the same information, but only for networks of constant density $\rho=0.25$ and size $n=100$ (b) and $n=500$ (c).
We also show the probability $\Pr(\mathrm{rNMI}(\mathcal{B}_p,\hat{\mathcal{B}})>0)$ that the inferred partition is correlated with the planted partition.
This a direct measurement of the fraction of detectable instances, which we compare against $\eta(\Delta;\rho,\beta,n)$.
It never reaches 0, because random fluctuations produce correlated partitions even when $\mathbb{P}=\mathbb{Q}$ (the rNMI corrects for the \emph{average} correlation).
If $\mathcal{L}>0$ were a necessary and sufficient condition for detectability, then $\eta(\Delta;\rho,\beta,n)$ and $\Pr(\mathrm{rNMI}>0|\Delta,\rho,\beta,n)$ would correspond perfectly.
But since $\mathcal{L}>0$ is only a \emph{necessary} condition, $\eta(\Delta)$ acts as an upper bound rather than an exact expression, i.e., $\Pr(\mathrm{rNMI}>0;\eta)$ can never be greater than $\eta(\Delta)$.

Two further observations must be made.
First, it is known that in the sparse two-blocks SSBM, the transition between the information-theoretically undetectable and detectable regions occurs on the so-called Kesten-Stigum (KS) bound---located at $\Delta=\pm q\sqrt{\rho/n}$ for finite-size instances (this is not generally true, but the equivalence holds when $q=2$ \cite{Mossel2013}).
Despite the fact that this bound was derived for infinite ensembles, it holds very well in the finite case, as shown in Figs.~\ref{fig:general_modular_graph_transition} (b) and \ref{fig:general_modular_graph_transition}~(c).
But the finite-size approach has the potential to be more precise.
Building upon the interpretation of $\avg{\mathcal{L}}$ as a measure of the average difficulty of the inference problem, we set a threshold $\avg{\mathcal{L}}=1/2n$ on the average detectability.
For this choice of threshold, the approximate hypersurface equation predicts a transition at
\begin{equation*}
    \Delta^* = \pm2\sqrt{\rho(1-\rho)/n}\;,
\end{equation*}
very close to the KS bound, but with a correction for nonvanishing densities.
Interestingly, one can motivate this choice of threshold with random matrix theory \cite{Young2014,Peixoto2013b,Nadakuditi2012} (see Appendix~\ref{appendix:connection_with_rmt} for details) or the theory of low-rank matrix estimation \cite{Lesieur2015}.
The uncorrected and corrected bounds are shown on Fig.~\ref{fig:general_modular_graph_transition}~(a).
The corrected bound is qualitatively accurate in all density regimes, unlike the KS bound.

Second, in asymptotic theories, the SBM is either said to be undetectable with overwhelming probability, or the converse.
The finite-size approach is more nuanced in the sense that it accounts for random fluctuations, which are also manifest in empirical results [see the curves $\Pr(\mathrm{rNMI}(\mathcal{B}_p,\hat{\mathcal{B}})>0)$].
While $\eta$--detectability is not perfect, as is argued above, it nonetheless goes through a smooth transition instead of an abrupt one.
This reflects the continuous nature of the finite-size transition.

%~~~~~~~~~~~~~~~~~~~~~~~~~~~~~~~~~~~~~~~~~~~~~~~~~~~~~~~~~~~~~~~~~~~~~~~~~~~~~~
\section{Conclusion}
\label{section:discussion}
%~~~~~~~~~~~~~~~~~~~~~~~~~~~~~~~~~~~~~~~~~~~~~~~~~~~~~~~~~~~~~~~~~~~~~~~~~~~~~~

Building upon ideas from statistical theory, we have developed a framework to study the information-theoretic detectability threshold of the finite-size SBM.
Our analysis relies on two different interpretations of the log-likelihood ratio $\mathcal{L}$ of the SBM and its equivalent random ensemble.
We have  used the rigorous interpretation of $\mathcal{L}$ to put a necessary condition on detectability.
We have then computed the distribution of $\mathcal{L}$, and proved that up to half of the instances of the finite-size SBM could be declared undetectable on the basis of this simple test alone.
We have further argued that the average of $\mathcal{L}$ could be interpreted as a proxy for the performance of an optimal inference algorithm (with possibly exponential running time).
This interpretation has proved to be fruitful; starting with a compact form for $\avg{\mathcal{L}}$, we have established the existence of a large equivalence class with respect to average detectability.
In Appendix~\ref{appendix:noisy_sbm}, we have shown that $\mathcal{L}$ can also be used to prove that, quite naturally, detectability decreases when the datasets are noisy.
Using a correspondence with the finite-size information-theoretic threshold (as well as with random matrix theory, see Appendix~\ref{appendix:connection_with_rmt}), we have presented numerical evidence that the hypersurface $\avg{\mathcal{L}}=1/2n$ separates detectable from undetectable instances in a special case of the SBM.

The unifying theme of this contribution has been the idea that $\avg{\mathcal{L}}$ quantifies both detectability and consistency in the finite-size SBM.
This interpretation leaves many questions open for future works.
Perhaps the most important of all: Can one determine the threshold within the framework of the theory itself, for general SBM?

A second important question pertains to sufficiency:
Can one modify the condition to make it necessary \emph{and} sufficient? Or is a completely different approach needed?
In asymptotic analyses of the limit, one can use different conditions to bound the limit from above and below, as is done in Ref.~\cite{Banks2016}.
Can a similar approach be fruitfully applied to finite instances?

In closing, let us mention a few of the many possible generalizations of the methods introduced.
First, it will be important to verify how our approach behaves in the limit $n\to\infty$.
How this limit is taken will matter.
In particular, we believe that our framework has much to say about the limit where $q\to\infty$, since it does not assume Poisson distributed degree, unlike other asymptotic theories of the limit.
Second, we see no major obstacle to a generalization of our methods to other generative models of networks with a mesoscopic structure.
This includes, for example, the consistency of graphons, a subject whose study has been recently undertaken \cite{Diao2016}.
Changing the null model from the equivalent random network ensemble to the configuration model \cite{Molloy1995,Newman2001} could even allow an extension to degree-corrected SBM \cite{Karrer2011}.

%==============================================================================
\section*{Acknowledgments}
%==============================================================================
We thank Charles Murphy and Guillaume St-Onge for useful discussions and comments.
This work has been supported by the Fonds de recherche du Qu\'ebec-Nature et technologies (J.-G.Y., P.D), the Conseil de recherches en sciences naturelles et en g\'enie du Canada (L.J.D.), the James S. McDonnell Foundation Postdoctoral Fellowship (L.H.-D.), and Grant No. DMS-1622390 from the National Science Foundation (L.H.-D.).
P.D. and E.L. are grateful to D.~C\^ot\'e (P.D., E.L) and P.~Mathieu (E.L.) for financial support.
{
\appendix

%~~~~~~~~~~~~~~~~~~~~~~~~~~~~~~~~~~~~~~~~~~~~~~~~~~~~~~~~~~~~~~~~~~~~~~~~~~~~~~
\section{Detectability and noise}
\label{appendix:noisy_sbm}
%~~~~~~~~~~~~~~~~~~~~~~~~~~~~~~~~~~~~~~~~~~~~~~~~~~~~~~~~~~~~~~~~~~~~~~~~~~~~~~

One almost never has a perfect knowledge of the structure of real networks.
The culprit can lie at the level of data collection, storage, transmission---or a combination of the above---, but the outcome is the same: Some edges are spurious and others are omitted \cite{Clauset2008}.
To model imperfect knowledge, we will suppose that instances of the SBM first go through a noisy channel where $T$ modifications---random edge removals or additions---are applied to the structure.
Only then are we asked to tell which of hypotheses  $\mathbb{P}$ and $\mathbb{Q}$ is the most likely.
It should be clear that it will be more difficult to separate the two hypotheses, since noise is not necessarily aligned with the planted partition.

We will approach the problem with the following \emph{universal perturbation process} (UPP):
At each step $t$ of this process, a new random edge is added with probability $c$; otherwise, a random edge is removed.
If a new edge must be added, then it is selected uniformly from the set of nonedges.
If an edge must be removed, then it is selected uniformly from the set of edges already present in the network.
This randomization step is then repeated $T$ times.
We call this process universal because one can map arbitrary perturbation patterns onto one or successive UPPs with different parameters $c$.

To prove that $\avg{\mathcal{L}}$ decreases as a result of any sufficiently long UPP, we will show that the total derivative
\begin{equation}
    \label{eq:appendix_total_derivative}
    \frac{d}{dt} \avg{\mathcal{L}}  = \sum_{r\leq s} \frac{\partial \avg{\mathcal{L}}}{\partial p_{rs}}  \frac{dp_{rs}(t)}{dt}
\end{equation}
is negative everywhere.
In so doing, we assume that the process can be approximated as a continuous one (both with regards to ``time'' $t$ and discrete quantities such as $m_{rs}$).
Admittedly, a more rigorous approach would be needed to settle the matter unequivocally, but we argue that the method presented in this appendix gives a good intuition for the problem.

Without specifying the dynamics, and using Eq.~\eqref{eq:avg_log_likelihood_entropy_difference}, one can compute
\begin{equation}
    \Pf{\avg{\mathcal{L}}}{p_{rs}} = \alpha_{rs}\log\left[\frac{p_{rs}}{\rho}\frac{1-\rho}{1-p_{rs}}\right] = \alpha_{rs} x_{rs}\;,
\end{equation}
where $x_{rs}$ is identical to Eq.~\eqref{eq:normalized_log_likelihood_x_rs}.
This leaves the $\dot{p}_{rs}(t)$ terms, whose expressions are determined by the perturbation dynamics.
For the UPP, the evolution of  $\{m_{rs}(t)\}_{r\leq s}$ is determined by the set of differential equations
\begin{equation}
    \label{eq:appendix_m_rs_dot}
    \!\dot{m}_{rs}(t)\! =\! - \frac{(1 - c)[m_{rs}(t)]}{\sum_{r\leq s}m_{rs}(t)}\! +\! \frac{c\ [m^{\max}_{rs} - m_{rs}(t)]}{m^{\max} - \sum_{r\leq s}m_{rs}(t)}.
\end{equation}
The first term accounts for edge removal events, which occur with probability $(1-c)$ and  involve edges that connect nodes in blocks ($B_r$, $B_s$) with probability $ m_{rs} / \sum m_{rs}(t) $.
A similar argument leads to the second term, which accounts for edge creation events.

Equation \eqref{eq:appendix_m_rs_dot} can be transformed into an equation for $\dot{p}_{rs}(t)$ by dividing through by $m^{\max}_{rs}$, and then using the definitions $p_{rs}(t)=m_{rs}(t)/m_{rs}^{\max}$ and $\rho(t)=\sum_{r\leq s} m_{rs}(t)/m^{\max}$.
We find
\begin{equation}
    \dot{p}_{rs}(t) =  \binom{n}{2}^{-1} \left [c\ \frac{1 - p_{rs}(t)}{1 - \rho(t) } - (1-c)\ \frac{p_{rs}(t)}{\rho (t)}\right]\;,
    \label{eq:appendix_p_rs_t}
\end{equation}
which, upon substitution in Eq.~\eqref{eq:appendix_total_derivative}, yields 
\begin{equation}
    \frac{d\avg{\mathcal{L}}}{dt}  = \Theta \sum_{r\leq s} \alpha_{rs}\log\left[\frac{f(p_{rs})}{f(\rho)}\right] \left [\frac{f(c) f(\rho)}{f(p_{rs})} - 1\right]\;,
\end{equation}
where $\Theta=[2(1-c)p_{rs}]/[\rho n(n-1)]$ is a nonnegative factor, and where we have defined $f(x)=x/(1-x)$.
It turns out that the sum is not only globally negative but that each term is also individually negative; i.e.,
\begin{equation}
    \label{eq:appendix_single_term}
    -\log\left[\frac{f(\rho)}{f(p_{rs})}\right] \left [\frac{f(c) f(\rho)}{f(p_{rs})} - 1\right] \leq 0\qquad \forall r\leq s.
\end{equation}
This comes about because the sign of the logarithm always matches that of the bracket.

To prove this statement, we treat five different cases and use the following identities repeatedly:
\begin{align}
    \label{eq:appendix_fxfy_ratio}
    &\frac{f(x)}{f(y)} < 1& &\!\!\!\!\!\!\implies\!\!\!\!\!\!& &x<y\;,&\\
    \label{eq:appendix_c_cond}
    &\frac{f(c)f(\rho)}{f(p_{rs})} > 1&  &\!\!\!\!\!\!\implies\!\!\!\!\!\!&  &c > \frac{p_{rs}(1-\rho)}{\rho(1-p_{rs}) + p_{rs}(1-\rho)}.&
\end{align}
The cases are:
\begin{enumerate}
    \item If $\rho=p_{rs}$: The logarithm equals 0 and the upper bound of Eq.~\eqref{eq:appendix_single_term} holds.
    \item If $p_{rs}<\rho$ and $c < 1/2$: The logarithm is positive [see Eq.~\eqref{eq:appendix_fxfy_ratio}].
    The bracket is also positive, since the inequality in Eq.~\eqref{eq:appendix_c_cond} can be rewritten as $(1-\rho)p_{rs}\leq \rho(1-p_{rs})$ using the fact that $c< 1/2$.
    This simplifies to $p_{rs} \leq \rho$, in line with our premise.
    \item If $p_{rs}<\rho$ and $c \geq 1/2$: The logarithm is positive.
    Using our premise, we conclude that $f(\rho)/f(p_{rs})>1$ and $f(c) \geq 1$.
    Therefore, $f(c)f(\rho)/f(p_{rs})>1$, i.e., the bracket is positive.
    \item If $p_{rs}>\rho$ and $c\leq 1/2$: The logarithm is negative. 
    Using our premise, we conclude that $f(\rho)/f(p_{rs})<1$ and $f(c) \leq 1$.
    Therefore, $f(c)f(\rho)/f(p_{rs})<1$, i.e., the bracket is negative.
    \item If $p_{rs}>\rho$ and $c>1/2$: The logarithm is negative.
    The bracket is also negative, since the converse of the inequality in Eq.~\eqref{eq:appendix_c_cond} can be rewritten as $(1-\rho)p_{rs}\geq \rho(1-p_{rs})$ using the fact that $c> 1/2$.
    This simplifies to $p_{rs} \geq \rho$, in line with our premise.
\end{enumerate}
This list covers all cases and therefore completes the proof that $d\avg{\mathcal{L}}/dt\leq 0$, i.e., that average detectability decreases as a result of the application of a UPP. \\

%~~~~~~~~~~~~~~~~~~~~~~~~~~~~~~~~~~~~~~~~~~~~~~~~~~~~~~~~~~~~~~~~~~~~~~~~~~~~~~
\section{Connection with random matrix theory}
\label{appendix:connection_with_rmt}
%~~~~~~~~~~~~~~~~~~~~~~~~~~~~~~~~~~~~~~~~~~~~~~~~~~~~~~~~~~~~~~~~~~~~~~~~~~~~~~
In Refs.~\cite{Nadakuditi2012,Peixoto2013b}  it is argued that SBM is not efficiently detectable when the extremal eigenvalues of the modularity matrix of its instances merge with the so-called ``continuous eigenvalue band.''
It is proved in Ref.~\cite{Nadakuditi2012} that this occurs when 
\begin{equation}
    \label{eq:appendix_poisson_limit}
    n(p_{\ins}-p_{\outs})=\pm\frac{1}{n}\sqrt{2n(p_{\ins}+p_{\outs})}\;,
\end{equation}
for the two-block SSBM with Poisson distributed degrees.
Furthermore, in this case, there is no so-called hard phase \cite{Mossel2013}, meaning that the above limit affords a comparison with the prediction if our information theoretic framework.

Since we are concerned with the finite case, let us first modify this result to account for binomial distributed degrees instead.
It turns out that the corrected condition is found by substituting the expectations of Poisson variables [in the right-hand-side of Eq.~\eqref{eq:appendix_poisson_limit}] by that of binomial variables.
This leads to
\begin{equation}
    \!\!\!(p_{\ins}-p_{\outs})\! =\! \pm \frac{1}{n}\sqrt{2n[p_{\ins}(1-p_{\ins}) + p_{\outs}(1-p_{\outs})]}\;,
\end{equation}
or, in terms of the natural parameters of the GMGM,
\begin{equation}
    \label{eq:appendix_rmt_prediction}
    \Delta^* =\pm \sqrt{\frac{4}{n-1}\rho(1-\rho)}\;.
\end{equation}
This equation bears a striking similarity with Eq.~\eqref{eq:explicit_hypersurface_GMG}, our approximate equation for curves of constant $\avg{\mathcal{L}}$.
In fact, for the two-block SSBM ($\beta\approx 1/2$), the latter reads
\begin{equation}
    \Delta^*=\pm\sqrt{8\lambda\rho(1-\rho)}\;.
\end{equation}
One obtains an exact equivalence between the two expressions by setting $\lambda = 1/2(n-1)\approx 1 /2n$.
The fact that modularity based spectral methods cannot infer a correlated partition if $\Delta\leq \Delta^*$ [Eq.~\eqref{eq:appendix_rmt_prediction}] can thus be understood as stemming from a lack of statistical evidence for the SBM.

%~~~~~~~~~~~~~~~~~~~~~~~~~~~~~~~~~~~~~~~~~~~~~~~~~~~~~~~~~~~~~~~~~~~~~~~~~~~~~~
\section{Detailed proofs}
\label{appendix:proof_details}
%~~~~~~~~~~~~~~~~~~~~~~~~~~~~~~~~~~~~~~~~~~~~~~~~~~~~~~~~~~~~~~~~~~~~~~~~~~~~~~
\subsection{Symmetries of the average detectability}
\label{appendix:proof_symmetries}

\begin{thm}[$\lambda$--preserving symmetries]
\label{thm:symmetries}
  All transformations $T(\bm{\alpha},\bm{P})$ of the parameter space of the SBM  that are (i) reversible, (ii) space-preserving, and (iii) valid at every point of the parameter space can be written as
  \begin{subequations}
  \label{eq_group:transformation_unconstrained_algebraic}
  \begin{align}
    \label{eq:transformation_unconstrained_algebraic_p}
    p_{rs} \mapsto  p_{rs}'&= \gamma_{rs} + (1-2\gamma_{rs})p_{\omega(r,s)}\;,\\
    \label{eq:transformation_unconstrained_algebraic_alpha}
    \alpha_{rs}\mapsto \alpha_{rs}'&= \alpha_{\pi(r,s)}\;,
  \end{align}
  \end{subequations}
  where $\gamma_{rs}\in\{0,1\}$ and where $\pi$ and $\omega$ are permutations that acts on the set $\{ (r,s)\,|\,1\leq r,\leq s\leq g\,\}$.
  Under the additional constraint that $\avg{\mathcal{L}(\bm{\alpha},\bm{P})}$ be preserved by $\{T\}$ and equal to $\lambda$, one must have
  \begin{equation*}
    \pi=\omega \quad\text{and}\quad \gamma_{rs}=\gamma\quad \forall (r,s)\;.
  \end{equation*}
\end{thm}

Let us first introduce new notations to clarify the proof of Theorem \ref{thm:symmetries}.
First, we define vectors $\ket{p}$ and $\ket{\alpha}$ whose entries are the $q^*=\binom{q}{2} + q$ entries of the upper triangle (and diagonal) of  $\bm{P}$  and $\bm{\alpha}$.
In this notation, we write the average density as $\braket{\alpha|p}$ and the average detectability as
\begin{align}
  \avg{\mathcal{L}(\bm{\alpha},\bm{P})} =\braket{\alpha|u(\bm{\alpha},\bm{P})}\;, \label{eq:braket_def_L}
\end{align}
where $\ket{u(\bm{\alpha},\bm{P})}$ is $q^*$--dimensional vector parametrized by $(\bm{\alpha},\bm{P})$, whose entries are given by
\begin{equation*}
  u_{rs}(\bm{\alpha}, \bm{P})=p_{rs}\log \frac{p_{rs}}{\langle  \alpha| p \rangle}+ (1-p_{rs})\log \frac{1-p_{rs}}{1-\langle  \alpha| p \rangle}\;.
\end{equation*}
We also introduce $\bm{\Pi}$ and $\bm{\Omega}$, two $q^*\times q^*$ permutation matrices such that $\bm{\Pi}\ket{\alpha}_{rs}=\alpha_{\pi(r,s)}$ and $\bm{\Omega}\ket{p}_{rs}=p_{\omega(r,s)}$, where $\ket{a}_{ij}$ is the element $(i,j)$ of vector $\ket{a}$.
In this notation, Eqs.~\eqref{eq_group:transformation_unconstrained_algebraic} are given by
  \begin{align*}
  \ket{\alpha} &\mapsto \ket{\alpha'} = \bm{\Pi} \ket{\alpha}\;,  \\
  \ket{p}      &\mapsto \ket{p'}=\bm{\Gamma}\ket{1} + (\bm{I}-2\bm{\Gamma})\bm{\Omega}\ket{p} \notag\\
               &\qquad \quad \equiv \bm{\Omega}\bm{\Gamma}'\ket{1} + \bm{\Omega}(\bm{I}-2\bm{\Gamma}')\ket{p} \;,
  \end{align*}
where $\bm{\Gamma}$ is a diagonal matrix with element $\gamma_{rs}$ on the diagonal, where $\bm{I}$ is the identity matrix, and where $\bm{\Gamma}'=\bm{\Omega}^{-1}\Gamma$ is also a diagonal matrix.

\begin{proof}
The proof of the first part of Theorem \ref{thm:symmetries} (form of the transformations) is given in the main text, see Sec.~\ref{subsubsection:symmetries_avg_detectability}. 

To prove the second part of the theorem (constrained transformations), we look for the subset of all transformations of the form shown in Eq.~\eqref{eq_group:transformation_unconstrained_algebraic} that also preserve $\avg{\mathcal{L}}$, i.e., transformations $T$ in $S_{q^{*}}\times B_{q^{*}}$ that map $(\bm{\alpha},\bm{P})$  to   $(\bm{\alpha}',\bm{P}')$ and that satisfy
\begin{equation*}
  \braket{\alpha|u(\bm{\alpha},\bm{P})}=
  \braket{\alpha'|u(\bm{\alpha}',\bm{P}')}\;.
\end{equation*}
It is easy to check that if $\bm{\Omega}=\bm{\Pi}$ and $\bm{\Gamma}=\gamma\bm{I}$ with $\gamma\in\{0,1\}$, then the average density  and the normalized log-likelihood are both preserved.
Therefore, if the transformations are of the proposed form, then $\lambda$ is preserved.

To complete the proof we must show that $\avg{\mathcal{L}}$ is conserved \emph{only if} $\bm{\Gamma}=\gamma\bm{I}$ and $\bm{\Omega}=\bm{\Pi}$.
First, we note that by the properties of the scalar product and permutation matrices, we have the following obvious symmetry 
\begin{equation*}
  \braket{\alpha|u} =\braket{\bm{\Pi}\alpha|\bm{\Pi}u},
\end{equation*}
which is valid for all permutation matrices $\bm{\Pi}$.
We use this symmetry to ``shift'' all permutation matrices to the second part of the scalar product representation of $\avg{\mathcal{L}}$, i.e., we write
\begin{equation*}
  \braket{\alpha|u} \mapsto \braket{\alpha'|u'} = \braket{\bm{\Pi}\alpha|u'} = \braket{\alpha|\bm{\Pi}^{-1}u'}\;.
\end{equation*}
Now, from Eq.~\eqref{eq:braket_def_L}, it is clear that we will have $\avg{\mathcal{L}(\bm{\alpha},\bm{P})}=\avg{\mathcal{L}(\bm{\alpha}',\bm{P}')}$ if and only if
\begin{equation}
\label{eq:scalar_product_condition}
  \braket{\alpha|u-\bm{\Pi}^{-1}u'} = 0\;,
\end{equation}
where $\ket{u'}:=\ket{u(\bm{\alpha}',\bm{P}')}$.
Since $\ket{u-\bm{\Pi}^{-1}u'}$ is analytic in $\bm{\alpha}$, we can expand it by using Taylor series;
this creates an infinite series of constraints that must all be satisfied.
In particular, the condition in Eq.~\eqref{eq:scalar_product_condition} will be satisfied only if
\begin{equation*}
  \ket{u-\bm{\Pi}^{-1}u'} = \ket{0}\;.
\end{equation*}
This is true if and only if, for all $(r,s)$, one has
\begin{multline}
 \label{eq:condition_by_term}
  p_{rs}\log \frac{p_{rs}}{\braket{\alpha|p}}+  (1-p_{rs})\log \frac{1-p_{rs}}{1-\braket{\alpha|p}}
  \\=\bar{p}_{rs}\log \frac{\bar{p}_{rs}}{\braket{\alpha|\bar{p}}}+
 (1-\bar{p}_{rs})\log \frac{1-\bar{p}_{rs}}{1-\braket{\alpha|\bar{p}}}\;,
\end{multline}
where $\ket{\bar{p}}=\bm{\Pi}^{-1}\ket{p'}$.
Here, $\ket{\bar{p}}$ is the transformed vector $\ket{p'}$, on which the inverse of permutation $\pi(r,s)$ is also applied.

Let us now suppose that $\bm{\alpha}$ tends to the point $\tilde{\bm{\alpha}}$, which is such that  $\tilde{\alpha}_{rs}=0$ for all $(r,s)$ except for $(r,s)=(a,b)$ (i.e., $\tilde{\alpha}_{ab}=1$).
In this limit, Eq.~\eqref{eq:condition_by_term} is 
trivially satisfied when $(r,s)=(a,b)$ but not otherwise. 
Let us suppose $(r,s)\neq (a,b)$ and expand the equation around $p_{ab}=\bar{p}_{ab}=\frac{1}{2}$.
From this second series expansion, one concludes that the equality is satisfied if either $\bar{p}_{ab} = p_{ab}$ or $\bar{p}_{ab}=1- p_{ab}$.
In both cases, the indices must match, which implies that $(a,b)=\pi^{-1}\circ \omega(a,b)$.
By repeating the same argument for all $(a,b)$, we conclude that $\omega=\pi$.
Thus, the map $T\, :(\bm{\alpha} ,\bm{P})\mapsto (\bm{\alpha}' ,\bm{P}')$ is a symmetry only if $\bm{\Pi} =\bm{\Omega}$.

This leaves the proof that $\bm{\Gamma}=\gamma\bm{I}$.
Let us, by contradiction, assume that $\gamma_{rs}$ differs from one set of indices to the other and define the sets $A$ and $B$ by
\begin{equation*}
  A=\{(r,s): \gamma_{rs}=0\} \quad \text{and} \quad B=\{(r,s): \gamma_{rs}=1\}\;.
\end{equation*}
Then one can write
\begin{equation}
  \rho = \braket{\alpha|p} = \avg{p}_A + \avg{p}_B\;,
\end{equation}
where $\avg{p}_X:=\sum_{(r,s) \in X} \alpha_{rs} p_{rs}.$
Returning to Eq.~\eqref{eq:condition_by_term} for $(r,s)\in A$ and using the newfound fact that $\bm{\Pi}=\bm{\Omega}$ which implies $\bar{p}_{rs}=\gamma_{rs}+(1-2\gamma_{rs})p_{rs}$ (no more permutations), we find 
\begin{multline*}
  p_{rs}\log \frac{p_{rs}}{\rho}+  (1-p_{rs})\log \frac{1-p_{rs}}{1-\rho}
  \\=p_{rs}\log \frac{p_{rs}}{\avg{p'}_A + \avg{p'}_B}+
 (1-p_{rs})\log \frac{1-p_{rs}}{1-\avg{p'}_A - \avg{p'}_B}\;.
\end{multline*}
This can only be true if $\rho=\avg{p'}_A + \avg{p'}_B$, i.e., if $A=\emptyset$ or $B=\emptyset$.
Therefore, $\gamma_{rs}=\gamma$ $\forall(r,s)$, with $\gamma\in\{0,1\}$.

\end{proof}

%~~~~~~~~~~~~~~~~~~~~~~~~~~~~~~~~~~~~~~~~~~~~~~~~~~~~~~~~~~~~~~~~~~~~~~~~~~~~~~
\subsection{Convexity of $\avg{\mathcal{L}}$}
\label{appendix:convexity_proof}
%~~~~~~~~~~~~~~~~~~~~~~~~~~~~~~~~~~~~~~~~~~~~~~~~~~~~~~~~~~~~~~~~~~~~~~~~~~~~~~
\begin{thm}
\label{thm:convexity}
  $\avg{\mathcal{L}(\bm{\alpha},\bm{P})}$ is convex with respect to $\bm{P}$.
\end{thm}
This property of $\avg{\mathcal{L}}$ is---perhaps surprisingly---not a consequence of the convexity of the KL divergence.
Instead, it follows from the log-sum inequality.
\begin{proof}
  We prove that $\avg{\mathcal{L}(\bm{\alpha},\bm{P})}$ is convex with respect to $\bm{P}$ by showing that it satisfies the convexity condition 
  \begin{multline}
    \label{eq:convexity_def}
    \avg{\mathcal{L}(\bm{\alpha},(1-t)\bm{P} + t\bm{Q})} \\ \leq (1-t)\avg{\mathcal{L}(\bm{\alpha},\bm{P})} + t\avg{\mathcal{L}(\bm{\alpha},\bm{Q})}\;,
  \end{multline}
  explicitly for all $t\in[0,1]$.
  Again, for the sake of clarity, we will use the notation developed in the previous section, and, in particular, write the density as $\rho=\braket{\alpha|p}$.
  We write each term on the left-hand-side of Eq.~\eqref{eq:convexity_def} as
  \begin{widetext}
  \begin{equation*}
    \alpha_{rs}\Bigg\{[(1-t)p_{rs} + tq_{rs} ]\log\frac{(1-t)p_{rs} + tq_{rs} }{(1-t)\braket{\alpha|p}+t\braket{\alpha|q}}+ [(1-t)(1-p_{rs})+t(1-q_{rs})]\log\frac{(1-t)(1-p_{rs})+t(1-q_{rs})}{(1-t)(1-\braket{\alpha|p})+t(1-\braket{\alpha|q})}  \Bigg\}
    \label{eq:term_explicit}
  \end{equation*}
  \end{widetext}
  It is easy to see that the log-sum inequality
  \begin{equation*}
  (a+\bar{a})\log\frac{a+\bar{a}}{b+\bar{b}}\leq a\log \frac{a}{b}+\bar{a}\log \frac{\bar{a}}{\bar{b}}
  \end{equation*}
  can be applied to both parts of Eq.~\eqref{eq:term_explicit} to separate terms by their coefficients $(1-t)$ and $t$.
  Repeating the same operation on all terms yields the inequality in Eq.~\eqref{eq:convexity_def}.
\end{proof}

% \bibliography{../reports/SBM}

\end{document}